\documentclass[
  prb,
  twocolumn,
  showpacs,
  amsmath,
  amssymb,
  superscriptaddress, 
]{revtex4}

\usepackage{bm}
\usepackage{graphicx}
\newcommand{\sign}{\mathop{\mathrm{sign}}}
\newcommand{\sh}{\mathop{\mathrm{sh}}}
\newcommand{\ch}{\mathop{\mathrm{ch}}}
\newcommand{\INT}{\mathrm{int}}
\newcommand{\trho}{\tilde{\rho}}
\begin{document}

\title{Relaxation in Luttinger liquids: Bose-Fermi duality}

\author{I.\ V.\ Protopopov}
\affiliation{
 Institut f\"ur Theorie der Kondensierten Materie and DFG Center for Functional
Nanostructures, Karlsruhe Institute of Technology, 76128 Karlsruhe, Germany
}
\affiliation{
 L.\ D.\ Landau Institute for Theoretical Physics RAS,
 119334 Moscow, Russia
}

\author{D.\ B.\ Gutman}
\affiliation{Department of Physics, Bar Ilan University, Ramat Gan 52900,
Israel }
\affiliation{
 Institut f\"ur Nanotechnologie, Karlsruhe Institute of Technology,
 76021 Karlsruhe, Germany
}

\author{A.\ D.\ Mirlin}

\affiliation{
 Institut f\"ur Nanotechnologie, Karlsruhe Institute of Technology,
 76021 Karlsruhe, Germany
}
\affiliation{
 Institut f\"ur Theorie der Kondensierten Materie and DFG Center for Functional
Nanostructures,
 Karlsruhe Institute of Technology, 76128 Karlsruhe, Germany
}

\affiliation{
 Petersburg Nuclear Physics Institute,  188300 St.~Petersburg, Russia.
}

\begin{abstract}

We explore the life time of excitations in a dispersive Luttinger liquid.
We perform a bosonization supplemented by a sequence of unitary transformations
that allows us to treat the problem in terms of weakly interacting
quasiparticles. The relaxation described by the resulting Hamiltonian is
analyzed by bosonic and (after a refermionization) by fermionic perturbation
theory. We show that the the fermionic and bosonic formulations of the problem
exhibit a remarkable strong-weak-coupling duality. Specifically, the fermionic
theory is characterized by a dimensionless coupling constant $\lambda= m^*l^2T$
and
the bosonic theory by $\lambda^{-1}$, where $1/m^*$ and $l$ characterize
the curvature of the fermionic and bosonic spectra, respectively, and $T$ is
the temperature. 

\end{abstract}


\maketitle

\section{Introduction}

Quantum kinetics in interacting one-dimensional (1D) systems is a subject of an
active experimental and theoretical investigation. There is a variety of
experimental realizations of 1D fermionic systems which include, in particular,
carbon nanotubes, semiconductor and metallic nanowires, as well as edge
states of quantum Hall systems and of other 2D topological insulator structures.
Further, cold atomic gases in optical traps can be used to engineer 1D
fermionic or bosonic systems with a tunable interaction. A light or microwaves
in waveguides with interaction mediated by two-level systems represent another
realization of a correlated 1D bosonic system.

A common and very powerful theoretical approach to interacting 1D systems is the
bosonization \cite{Stone_book,Delft,Gogolin,Giamarchi}. When the
spectral  curvature and backscattering processes are neglected, bosonization
maps the problem of interacting fermions (known as Tomonaga-Luttinger model) to
the Luttinger liquid theory of free bosons (plasmons). A mapping to the
Luttinger liquid is also obtained if one starts from the problem of bosons with
repulsion. If the interaction is considered as momentum-independent (i.e. local
in the coordinate space), the spectrum of Luttinger-liquid bosonic excitations
is linear, and by virtue of refermionization the problem is equivalent to that
of free fermions. In brief, when the momentum dispersions of excitations are
neglected, the fermionic and bosonic 1D problems are equivalent, and the
interaction can be completely eliminated.  

The problem becomes much more complex when both the spectral curvature of
constituent particles and the momentum dependence of the interaction are
retained. This leads (apart from some special cases) to a violation of
integrability of the theory. While the corresponding corrections to the
Luttinger liquid theory are irrelevant in the renormalization-group (RG) sense,
they are very important from the physical point of view. Specifically, they
establish a finite relaxation rate of excitations when the system is at non-zero
temperature or away from equilibrium. 

Several recent works addressed various aspects of relaxation in 1D problems. In
Refs.~\onlinecite{Lunde07,Khodas2007,imambekov09,imambekov11} a perturbative
analysis of three-particle scattering in a model of weakly interacting fermions
with a spectral curvature (inverse mass) $m^{-1}$ was performed. It was found,
in particular, that in the case of spinless fermions the intrabranch scattering
processes (RRL $\to$ RRL and RLL $\to$ RLL, where R and L denote right- and
left-movers, respectively) induce a scattering rate of an excitation with
momentum $k$ that scales as $(k-k_F)^8/m^3$ at zero temperature and $(k-k_F)^6
T/m^2$ at sufficiently high temperature $T$, where $k_F$ is the Fermi momentum.
These results were generalized to the case of Coulomb interaction in
Refs.~\onlinecite{micklitz11,ristivojevic13}. 

The opposite limiting case of Luttinger liquid with interaction parameter $K\ll
1$ (describing, in particular, fermions with very strong repulsive interaction,
when the system can be viewed as ``almost a Wigner crystal'') was
considered in Refs.~\onlinecite{apostolov13,Lin2013}. The authors of these works
analyzed the decay rate of bosonic excitations in such systems and found the
decay rate scaling as $T^5$. 

The goal of this work is to study systematically relaxation in dispersive
Luttinger liquids in the whole space of parameters. We start from the
interacting fermionic problem, then bosonize it and perform a unitary
transformation \cite{PGM2013} (that extends the one originally
introduced in Ref.~\onlinecite{Matis_Lieb1965}; see also
Refs.~\onlinecite{Rozhkov,imambekov11,MatveevFurusaki2013}) which allows one to
eliminate a major part of the interaction. In particular, in this way the two
chiral branches get decoupled up to the third order in density fluctuations.
In Ref.~\onlinecite{PGM2013} this formalism was employed to develop a
formalism of kinetic equation for fermionic quasiparticles. A focus in that
work was on a not too long time scale where the collisions between
quasiparticles can be neglected. Here we use the theory resulting
from the above unitary transformation to find the relaxation rate of
excitations. We perform both fermionic and bosonic analysis of this theory,
evaluate the corresponding relaxation rates, and determine regions of the
parameter space where each of the approaches is applicable. This allows us to
establish a remarkable picture of Fermi-Bose duality in dispersive interacting
1D systems.  

The structure of this article is as follows. In Sec.~\ref{sec:Model} we
introduce a model of a generic dispersive Luttinger liquid in terms of fermions
with a spectral curvature $1/m$ and an interaction with an arbitrary
strength and a radius $l_{\rm int}$. The fact that both parameters $1/m$
and $l_{\rm int}$ are non-zero makes the problem non-integrable. We
perform a bosonization of this model supplemented by a unitary transformation
that maps it onto a problem of weakly interacting bosonic quasiparticles. In
Sec.~\ref{sec:Bosons} we calculate the lifetime of bosonic excitations within
this theory. In Sec.~\ref{sec:Fermions} we refermionize the theory obtained in
Sec.~\ref{sec:Model} and explore the relaxation of fermionic excitations. For
this purpose, we calculate the contributions to this relaxation rate from both
inter-branch and intra-branch scattering processes. Finally, in
Sec.~\ref{sec:Duality} we collect and analyze the obtained results and determine
the behavior of the relaxation rate in the whole parameter space. We
show that the parameter space is subdivided in the ``fermionic'' and
``bosonic'' domains with a dimensionless control parameter $m^*l^2 T$, where
the effective mass $m^*$ and the plasmon dispersion length $l$ are expressed in
terms of the bare parameters $m$, $l_{\rm int}$ and the Luttinger-liquid
constant $K_0$. The emerging picture has a character of Fermi-Bose weak-strong
coupling duality. We close the paper by summarizing our results and discussing
prospects for future research in Sec.~\ref{sec:Summary}.

\section{Dispersive Luttinger liquids}
\label{sec:Model}

In this section we introduce the Hamiltonian of our model and
formulate basic questions to be addressed in the paper.  
We then perform a sequence of unitary transformations  bringing the theory to
a form amenable to a perturbative treatment and highlighting the duality between
fermionic and bosonic descriptions of the dispersive Luttinger liquids.

\subsection{The model}

Our starting point is the Hamiltonian  of a generic ``dispersive'' Luttinger 
liquid comprised of  (spinless) right- and left-moving fermions 
(created and annihilated by operators $\psi^+_{\eta}(x)$, $\psi_{\eta}(x)$ with
$\eta=R, L$;  occasionally, we also use 
the notation $\eta=\pm 1$)  with curved single-particle spectrum
$\epsilon_\eta(k)=\eta k v_F+k^2/2m$ interacting via  a
generic finite-range density-density interaction $g(x)$
\begin{multline}
 H=\sum_{\eta}\int dx \psi^+_\eta(x)\left(-i\eta v_F\partial_x-\frac{1}{2m}\partial_x^2\right)\psi_\eta(x)\\
 +\frac{1}{2}\int dx dx' g(x-x')\rho(x)\rho(x')\,.
 \label{sec2:Hamiltonian}
\end{multline}
Here $\rho(x)=\rho_{R}(x)+\rho_{L}(x)$. We characterize the interaction $g(x)$
by its strength at zero momentum $g_0$ and its radius  
$l_{\INT}\gtrsim 1/(m v_F)$ so that in momentum space 
\begin{equation}
 g_{q}=g_0\left(1-q^2l_{\INT}^2\right)\,,\qquad ql_{\INT}\ll 1\,.
\end{equation}
We are interested in the properties of our model at low momenta $q\ll l_{\mathrm
int}\lesssim p_F\equiv m v_F$ and  energies $\epsilon\ll p_F v_F$.   
 
In the subsequent consideration  we will neglect  the processes
changing the total number of fermions $N_\eta$ (counted from its value in the
ground state) within each chiral branch. 
They are absent in our model Hamiltonian (\ref{sec2:Hamiltonian}) but are of
course present in any real 1D system. These processes play a crucial role  
in the ultimate equilibration between branches in the Luttinger liquid
\cite{Lunde07,Matveev2013,Dmitriev2012} 
but show up only at exponentially large time scales $\propto \exp(E_F/T)$ and
are completely irrelevant for the physics discussed in this work.  
Accordingly, from now on we consider our system in the sector characterized by
$N_R=N_L=0$ and set the zero Fourier components of the densities
$\rho_\eta(x)$ to zero.  

The standard Tomonaga-Luttinger (TL) is the extreme low-energy limit  of the
Hamiltonian (\ref{sec2:Hamiltonian}) 
 corresponding to linear fermionic spectrum ($m=\infty$) and point-like
interaction $g(x)=g_0\delta(x)$. From the RG perspective, contributions that
are neglected within this approximation are irrelevant perturbations. 
Specifically, when setting $m=\infty$, one drops an
irrelevant perturbation of scaling dimension $3$, while discarding  
the momentum dependence of the interaction is equivalent [for a finite-range
$g(x)$] to the neglect of even weaker 
 perturbation of scaling dimension $4$. The bosonization approach
\cite{Stone_book, Delft, Giamarchi, Gogolin} allows one to map the TL
Hamiltonian onto free dispersionless  bosons, which in turn are equivalent via
refermionazation \cite{Matis_Lieb1965, Rozhkov, imambekov09, imambekov11} to
free fermions. Thus, the TL model can be equally well treated in bosonic and
fermionic (after the identification of the proper fermionic modes) languages. 
This fact is related to the conformal invariance of the TL Hamiltonian. 

Despite the great success of the TL model in the description of thermodynamic
properties of 1D interacting fermions, it is now known  that the irrelevant
perturbations it neglects can have strong impact on the dynamical response  
of the system. For example, the fermionic curvature translates upon bosonization
into a cubic interaction of the density fluctuations\cite{schick68}
\begin{equation}
-\frac{1}{2m}\int dx\psi^+_{\eta}(x)\partial_x^2\psi_\eta(x)\\=\frac{2\pi^2}{3m}\int dx\rho_\eta^3(x)\,.
\end{equation}
Although irrelevant in the RG sense, this perturbation acts 
for the case of a short-range interaction $g(x)$ (or just for free fermions), 
on a highly degenerate linear bosonic  spectrum, so that the corresponding
perturbation theory suffers from strong divergences. As a consequence, the 
formally irrelevant perturbation alters dramatically the behavior 
of, e.g., single particle spectral weight $A(k, \epsilon)$ in the immediate
vicinity of the single-particle mass shell\cite{imambekov09, imambekov11}. 

While extremely nontrivial in the bosonic representation, the problem of
Luttinger liquid with finite fermionic mass can be elegantly addressed  via the
introduction of the proper fermionic quasiparticles (refermionization) 
\cite{Rozhkov, imambekov09, imambekov11}. Thus the fermionic curvature breaks
the symmetry between fermionic and bosonic languages present in the TL model in
favor of fermions.  

Conversely, for fermions with linear spectrum  ($m=\infty$) the Hamiltonian
(\ref{sec2:Hamiltonian}) describes (after bosonization) free bosons with the
dispersion relation
\begin{equation}
 \omega_q= u_q |q|\,, \qquad u_q=v_F\left(1+\frac{g_q}{\pi v_F}\right)^{1/2}\equiv \frac{v_F}{K_q}\,.
\label{sec2:uq}
\end{equation}
At small momenta the  boson velocity $u_q$ is given by
\begin{eqnarray}
 u_q=u_0\left(1-l^2 q^2\right)\,,\\
 l^2=\frac{1}{2}(1-K_0^2)l_{\INT}^2\,,
\end{eqnarray}
where $K_0$ is the zero-momentum limit of the Luttinger-liquid parameter $K_q$
introduced in Eq.~(\ref{sec2:uq}). 
Finite interaction radius $l_{\INT}$ leads thus to the appearance of dispersion in the bosonic spectrum. 
For long-range interactions the Wigner-crystal-type correlations proliferate and
{\it bosonic} excitations are 
stable against perturbations caused by the curvature of fermionic spectrum\cite{Lin2013} in a wide range of 
energies. Note that upon refermionization, the curvature of bosonic spectrum translates  into  
an interaction between fermionic quasiparticles \cite{imambekov09, PGM2013}. 

The consideration just presented raises the fundamental question: What are the
proper degrees of freedom for the description of a generic dispersive 
Luttinger liquid having both curved  fermionic and bosonic spectra? We observe a
remarkable duality  between the fermionic and bosonic description of the
problem: curved single-particle spectrum for the excitations of  one type
(fermions or bosons) introduces the interaction between the excitations of  the
other type. The importance of this interaction for the dynamics of the particles
of the second type is determined in turn by the curvature of their own spectrum.
Accordingly, we expect that in a generic dispersive Luttinger liquid the
particles with  the most curved spectrum are the most long-living and well
defined excitations. For a finite-range interaction $g(x)$, the nonlinear
corrections to the bosonic and fermionic excitation spectra scale
differently with momentum. Specifically, at sufficiently large momenta the
bosonic correction $\delta\omega_q\propto v_F l_{\rm int}^2 q^3$ dominates
over the fermionic correction $\delta \xi_k\propto k^2/m$,  while at small
momenta the situation is reverse. Thus, one can expect that if the
characteristic energy scale of the problem (say, temperature $T$) exceeds
$T_0\propto 1/m l_{\rm int}^2$, the bosonic language gives the proper
description of relaxation in the system, while at smaller energies the fermionic
language becomes appropriate.

In the rest of the paper we explore the life time of bosonic and fermionic
excitations in the dispersive Luttinger liquid. 
To be definite, we consider the Luttinger liquid  at finite temperature $T$ and
study the decay rate $1/\tau_\epsilon(T)$ of a right-moving  excitation (boson
or fermion) injected into the system at energy $\epsilon\gtrsim T$.
In agreement with the qualitative consideration presented above, we find that at
sufficiently large $T$, $\epsilon$ the perturbatively obtained life time of
bosonic excitations is much longer than that of fermionic ones. In this regime,
the bosonic perturbation theory is justified and yields the correct relaxation
rate. The situation is reversed at low $T$, $\epsilon$: in this case the
fermionic calculation of the relaxation rate becomes controllable, and the
fermionic quasiparticles are proper excitations. The correspondence between the
two approaches can be viewed as an example of a strong-weak coupling duality in
physics.

\subsection{Unitary transformations}

In this subsection we seek for the representation of the dispersive Luttinger
liquid in terms of weakly interacting 
quasiparticles. The original fermions interact strongly. 
The RG classification of various terms in the Hamiltonian (\ref{sec2:Hamiltonian}) 
suggests  that in order to reduce the interaction we need first to get
rid of the density-density interaction between right- and left- moving fermions.
The natural way to achieve this goal is bosonization. Thus  we bosonize 
the Hamiltonian  (\ref{sec2:Hamiltonian}) and arrive at
\begin{multline}
 H=\sum_{\eta}\int dx :\left(\pi v_F\rho_\eta^2(x)+\frac{2\pi^2} {3m}\rho_\eta^3(x)\right):_B\\
 +\frac{1}{2}\int dx dx' g(x-x'):\rho(x)\rho(x'):_B\,.
 \label{sec2:HamiltonianBoson0}
\end{multline}
Here $::_B$ stands for normal ordering with respect to the bosonic modes
(Fourier components of $\rho_\eta(x)$).

The density-density coupling between left- and right- chiral sectors can be eliminated by a unitary 
transformation of bosonic operators\cite{Rozhkov}
\begin{eqnarray}
&& \rho_\eta(x)=U_2^+\trho_\eta(x)U_2 \,,\\
&& U_2=\exp\left[\frac{2\pi}{L}\sum_{q\neq 0}\frac{\kappa_q}{q}\trho_{R,
q}\trho_{L,-q}\right].
\label{sec2:U2}
\end{eqnarray}
Here the function  $\kappa_q$ is to be chosen from the requirement that
interbranch density-density interaction is absent in the transformed Hamiltonian
and we have assumed that the fermions resides on a circle of circumference $L$.
Taking into account the commutation relations of the density components
$\left[\rho_{\eta, q}, \rho_{\eta, -q}\right]=\eta L q/2\pi$, we see that $U_2$
generates the Bogoliubov transformation:
\begin{eqnarray}
&&
\rho_{R, q}=\cosh \kappa_q \trho_{R, q}-\sinh\kappa_q \trho_{L, q}\,, \\&& 
\rho_{L, q}=-\sinh \kappa_q \trho_{R,q} +\cosh\kappa_q \trho_{L, q}\,.
\end{eqnarray}
The decoupling of the chiral sectors of the theory to quadratic order in densities fixes now  the rotation angle  
\begin{equation}
\tanh 2\kappa_q= g_q/(2\pi v_F+g_q)\,.
\end{equation}
At small momenta we obtain
\begin{equation}
 \kappa_q=\kappa_0-\frac12 l^2 q^2\,.
\end{equation}

The unitary transformation $U_2$ fully solves the model of Luttinger liquid with
linear fermionic dispersion. In the generic situation that we are considering
this is no longer the case. In terms of the new density operators 
$\tilde{\rho}_\eta$ the Hamiltonian reads
\begin{eqnarray}
 H &=& (\pi/L)\sum_{q}u_q:\left(
\tilde{\rho}_{R, q}\tilde{\rho}_{R, -q}+
\trho_{L, q}\trho_{L, -q}\right):_B \nonumber \\
&+& \frac{1}{L^2}\sum_{{\bf q}}
 \left[\Gamma^{B, RRR}_{\bf q} :(\trho_{R, q_1}\trho_{R, q_2}\trho_{R,
q_3}+R\rightarrow L):_B \right. \nonumber \\&+& \left.
      \Gamma^{B, RRL}_{\bf q} :(\trho_{R, q_1}\trho_{R, q_2}\trho_{L, q_3}+R\leftrightarrow L):_B
\right]\,.
\label{sec2:HamiltonianBoson1}
\end{eqnarray}
Here ${\bf q}$ stands for $\{q_1,q_2,q_3\}$, the bosonic velocity $u_q$ was
defined in Eq. (\ref{sec2:uq}), and the three-boson interaction vertices are
given by 
\begin{eqnarray}
\Gamma^{B, RRR}_{\bf q}=\frac{2\pi^2}{3m }\left[\ch \kappa_1\ch \kappa_2 \ch
\kappa_3-\sh \kappa_1\sh \kappa_2\sh \kappa_3\right],
\label{sec2:GammaB_RRR_general}
\\
\Gamma^{B, RRL}_{\bf q}=\frac{2\pi^2}{m }\left[\sh \kappa_1\sh \kappa_2\ch
\kappa_3 -\ch \kappa_1\ch\kappa_2\sh \kappa_3\right],
\label{sec2:GammaB_RRL_general}
\end{eqnarray}
with $\kappa_i\equiv \kappa_{q_i}$. In Eqs. (\ref{sec2:GammaB_RRR_general}) and 
 (\ref{sec2:GammaB_RRL_general}) we have suppressed the Kronecker symbol
$\delta_{q_1+q_2+q_3, 0}$  
expressing the momentum conservation.

The vertices $\Gamma^{B, RRR}_{\bf q}$ and $\Gamma^{B, RRL}_{\bf q}$ can be
expanded at small momenta $q_i$:
\begin{eqnarray}
 \Gamma^{B, RRR}_{\bf q}&=&\frac{2\pi^2}{3 m^* }\left(1-\frac{\alpha l^2}{2}
(q_1^2+q_2^2+q_3^2)\right)\,,
 \label{sec2:GammaB_RRR}
 \\
 \Gamma^{B, RRL}_{\bf q}&=&-\frac{2\pi^2\alpha}{m^* }\left(1+\frac{1}{2} l^2
(q_1^2+q_2^2)-\frac{ l^2 q_3^2}{2\alpha} \right)\,.
\label{sec2:GammaB_RRL}
\end{eqnarray}
Here we have introduced the renormalized mass
\begin{equation}
 \frac{1}{m^*}=\frac{3+K_0^2}{4\sqrt{K_0}}\frac1m
 \label{sec2:m*}
\end{equation}
and a dimensionless parameter $\alpha$ characterizing the interaction strength,
\begin{equation}
 \alpha=\frac{1-K_0^2}{3+K_0^2}.
\end{equation}

From the RG prospective  the Hamiltonian (\ref{sec2:HamiltonianBoson1}) is a
Hamiltonian of free bosons with linear spectrum $\omega=u_0 q$  perturbed by i)
terms of scaling 
dimension 3 due to cubic interaction of bosons $\Gamma^{B, RRR}_{\bf q=0}$ and
$\Gamma^{B, RRL}_{\bf q=0}$; ii) a perturbation of dimension 4 originating from
the curvature of bosonic spectrum; iii) various terms of higher scaling
dimensions. At low energies it is natural to begin by taking care 
of most relevant perturbations, namely, the cubic bosonic couplings with
vertices approximated by their value at zero momentum. We first include the term
$\Gamma^{B, RRR}_{\bf q=0}$ that couples the bosons on the same highly
degenerate branch.
The resulting Hamiltonian 
\begin{multline}
 H=(\pi/L)\sum_{q}u_0\left(
:\tilde{\rho}_{R, q}\tilde{\rho}_{R, -q}:_B+
:\trho_{L, q}\trho_{L, -q}:_B\right)\\
+\frac{1}{L^2}\Gamma^{B, RRR}_{\bf q=0}\sum_{{\bf q}}
(:\trho_{R, q_1}\trho_{R, q_2}\trho_{R, q_3}:_B+R\rightarrow L) 
\label{sec2:HamiltonianCurvatureBosons}
\end{multline}
is just a bosonized version of a Hamiltonian of  non-interacting fermions with
the Fermi velocity $u_0$ and the spectral curvature  $1/m^*$,
\begin{multline}
 H=\sum_{\eta}\int dx \tilde{\psi}^+_\eta(x)\left(-i\eta u_0\partial_x-\frac{1}{2m^*}\partial_x^2\right)\tilde{\psi}_\eta(x)\,.
 \label{sec2:HamiltonianCurvature}
\end{multline}
The Hamiltonian (\ref{sec2:HamiltonianCurvature}) is an effective Hamiltonian of
the system at lowest  energies. It is worth mentioning
that the expressions (\ref{sec2:uq}) and (\ref{sec2:m*}) for the
coefficients $u_0$ and $m^*$ in the effective Hamiltonian
(\ref{sec2:HamiltonianCurvature}) are not exact because of corrections 
from the neglected perturbations and originating at the ultraviolet scale.
These corrections are small in the limit of long-range interaction, $l_{\rm
int}p_F \gg 1$  but generate non-trivial renormalization factors of order
unity for $l_{\rm int}p_F \sim 1$; see also Eq.~(\ref{sec2:deltaH2})
and a discussion following it. The exact values $u_0$ and $m^*$ can be
related to the thermodynamic characteristics of the system\cite{imambekov11}. 

We now reintroduce in the bosonized
Hamiltonian Eq.~(\ref{sec2:HamiltonianCurvature}) the terms describing the
curvature of the bosnic spectrum as well as the inter-branch cubic
couplings $\Gamma^{B, RRL}_{\bf q}$. 
Remarkably, it turns out to be possible\cite{PGM2013} to get rid of the
$\Gamma^{B, RRL}_{\bf q}$ terms. Indeed, it is easy to see that vertex
$\Gamma_{\bf q}^{B, RRL}$ does not describe a real scattering of bosons due to
impossibility to fulfill the momentum end energy conservation. It is thus
possible to design a unitary transformation $U_3$ 
eliminating the $\Gamma_{\bf q}^{B, RRL}$ coupling\cite{Remark_Zakharov}
\begin{equation}
\trho_R(x)=U_3^+R(x)U_3\,,\qquad \trho_L(x)=U_3^+L(x)U_3\,.
\end{equation}

The analogy with the Bogoliubov transformation $U_2$ suggests the
following ansatz for $U_3$:
\begin{eqnarray}
U_3 &\equiv& \exp[\Omega_3] \nonumber \\ &=& \exp\left\{\frac1{L^2} \sum_{{\bf
q}}[f_{\bf q} R_{q_1} R_{q_2} R_{q_3} - (L\leftrightarrow R)]\right\}.
\label{sec2:U3}
\end{eqnarray}
Performing a perturbative expansion of $U_3$, we obtain 
\begin{eqnarray}
 \trho_{R, q}&=& R_q-\left[\Omega_3, R_q\right]+\frac{1}{2}\left[\Omega_3,
\left[\Omega_3, R_q\right]\right]+{\mathrm
O}\left(\rho^4/p_F^3\right),\nonumber\\
 \trho_{L, q}&=&L_q-\left[\Omega_3, L_q\right]+\frac{1}{2}\left[\Omega_3,
\left[\Omega_3, L_q\right]\right]+{\mathrm
O}\left(\rho^4/p_F^3\right).\nonumber \\ &&
 \label{sec2:DensityExpansion}
\end{eqnarray}
In Eq. (\ref{sec2:DensityExpansion}) we  kept terms up to the third order
in densities which are required to compute the Hamiltonian in new variables up
to the fourth order. 

Neglecting  the third order terms in Eqs.(\ref{sec2:DensityExpansion}), substituting the resulting expansions into Hamiltonian (\ref{sec2:HamiltonianBoson1}) and demanding that the left- and right- sectors are decoupled at cubic order one finds
\cite{PGM2013}
\begin{equation}
 f_{\bf q}=\frac{\Gamma^{B, RRL}_{\bf q}}{u_{q_1} q_1+u_{q_2} q_2-u_{q_3} q_3}\,.
\label{sec2:f}
 \end{equation}
Note that the  impossibility to conserve momentum and energy in a scattering event involving two right bosons and one left boson guarantees that the energy denominator in 
(\ref{sec2:f}) is non-zero. The behavior of  $f_{\bf q}$ at small momenta can be
easily inferred from (\ref{sec2:f}) and (\ref{sec2:GammaB_RRL}):  
\begin{equation}
 f_{\bf q}=\frac{\pi^2\alpha}{m^* u_0 q_3}\left[1+\frac54 l^2(q_1^2+q_2^2)+ \frac{l^2}{4}\left(1-\frac{2}{\alpha}\right) q_3^2 \right]\,.
\end{equation}

Retaining now the third-oder terms in (\ref{sec2:DensityExpansion}), one can
recast the  Hamiltonian (\ref{sec2:HamiltonianBoson1}) into the form 
\begin{eqnarray}
 H &=& (\pi/L)\sum_{q}u_q\left(
:R_qR_{-q}:_B+:L_qL_{-q}:_B\right) \nonumber \\
&+& \frac{1}{L^2}\sum_{{\bf q}}
 \Gamma^{B, RRR}_{\bf q} (:R_{q_1}R_{q_2}R_{q_3}:_B+R\rightarrow L) \nonumber
\\ &+&
 \frac{1}{L^3}\sum_{\bf q} \Gamma^{B, RRRR}_{\bf
q}\left[R_{q_1}R_{q_2}R_{q_3}R_{q_4}+R\rightarrow L\right]
 \nonumber \\  &+&
 \frac{1}{L^3}\sum_{\bf q} \Gamma^{B, RRRL}_{\bf
q}\left[R_{q_1}R_{q_2}R_{q_3}L_{q_4}+R\leftrightarrow L\right] \nonumber \\
 &+& \frac{1}{L^3}\sum_{\bf q} \Gamma^{B, RRLL}_{\bf
q}R_{q_1}R_{q_2}L_{q_3}L_{q_4}+{\mathrm O}(\rho^5)
\,.
\label{sec2:HamiltonianBoson2}
\end{eqnarray}
In the last three sums ${\bf q}$ stands for  $\{q_1,q_2,q_3, q_4\}$. The couplings  $\Gamma^{B, RRRR}_{\bf q}$, $\Gamma^{B, RRRL}_{\bf q}$ and $\Gamma^{B, RRLL}_{\bf q}$ 
are symmetric functions of momenta corresponding to the density components of the same chirality. The full expressions for them  are cumbersome and we do not present them here (see Appendix \ref{app:HamiltonianBosons} for details). We will discuss their relevant properties when appropriate. 

We concentrate now on the general structure of
Eq.~(\ref{sec2:HamiltonianBoson2}) and notice several points that allow us to
simplify the Hamiltonian.
First, we neglect the ${\mathrm O}(\rho^5)$
corrections in the Hamiltonian (\ref{sec2:HamiltonianBoson2}) that have formal
smallness of $\rho^3/p_F^3$ as compared to the quadratic part. Indeed, our
calculation of the relaxation rates below shows that dominant contributions
originate from ${\mathrm O}(\rho^4)$ terms, so that there is no need to keep
terms of higher orders.

Second, we note the absence of normal ordering in the last three terms of the
Hamiltonian (\ref{sec2:HamiltonianBoson2}). Performing the bosonic normal
ordering, one generates various quadratic couplings of densities. For example,
the normal ordering of $\Gamma^{B, RRRL}_{\bf q}$ coupling generates the
contribution
\begin{equation}
 \delta H^{(2)}_{RL}\propto \sum_{p, q}\Theta(p)
 p\left(\Gamma^{RRRL}_{p, -p, q, -q}+\Gamma^{RRRL}_{p, -p, -q, q}\right)R_qL_{-q}\,.
 \label{sec2:deltaH2}
\end{equation}
The interaction of right and left movers described by (\ref{sec2:deltaH2}) is
finite at zero momentum. Its precise value is determined by the  behavior of 
$\Gamma^{RRRL}_{\bf q}$ at large momenta. A quick estimate shows that  these
corrections are small (in the parameter  $1/p_F l_{\INT}$) compared to the
density-density interaction in the initial Hamiltonian (\ref{sec2:Hamiltonian})
as long the interaction radius $l_{\INT}$ is large. One can get rid of the
generated quadratic couplings by a suitable modification of the unitary
transformation $U_2$. Obviously, (\ref{sec2:deltaH2}) and similar terms arising
from normal ordering are responsible for the renormalization of the Luttinger
parameter $K_0$ and other parameters of the effective theory coming from the
residual interactions at large energies. This renormalization is small for 
$l_{\INT}p_F\gg 1$ and becomes of order unity at $l_{\INT}\sim\lambda_F$. We
assume from now on that the transformation $U_2$ was suitably adjusted and omit
the terms arising from the bosonic normal ordering. 

The third simplification is as follows.
The vertex $\Gamma^{B, RRRR}_{\bf q}$ is non-singular at zero momentum 
\begin{equation}
 \Gamma^{B, RRRR}_{\bf q}=-\frac{\pi^3\alpha^2}{2 u_0 {m^*}^2 L^3}\left[
1-\frac{8-23\alpha}{24\alpha}l^2\sum_{i=1}^4 q_i^2
\right].
\end{equation}
Translated  to the fermionic representation it gives rise to i) a correction to
the fermionic spectrum $\delta \xi_k\propto k^3/m p_F$; ii) a small correction
to the density-density interaction of fermions of the same chirality $\propto
q^2R_{q}R_{-q}/p_F^2$; iii) various terms of higher scaling dimension. Since we
are interested in phenomena at energies much less than the Fermi energy, we can
neglect these corrections altogether.  

The last remark to be made on Eq. (\ref{sec2:HamiltonianBoson2}) is that
the momentum and energy conservation does no allow scattering processes
involving two right and two left bosons. Accordingly, the $\Gamma^{B, RRLL}_{\bf
q}$ coupling does not lead to real bosonic transitions in the first order of
perturbation theory and can be removed by a unitary transformation $U_4$
analogous to $U_3$. Apart from a modification of the ${\mathrm O}(\rho^5)$ terms
(which we neglect anyway), the elimination of  $\Gamma^{B,
RRLL}_{\bf q}$ coupling from the Hamiltonian (\ref{sec2:HamiltonianBoson2}) is
the only effect of transformation $U_4$.

We are now ready to summarize our findings on the structure of the
Hamiltonian. Once the unitary transformations are performed and terms
that give subdominant contributions to the relaxation are neglected, the
Hamiltonian of a dispersive Luttinger liquid can be presented as
\begin{eqnarray}
 H &=& (\pi/L)\sum_{q}u_q:\left(
R_qR_{-q}+L_qL_{-q}\right):_B \nonumber \\
&+& \frac{1}{L^2}\sum_{{\bf q}}
 \Gamma^{B, RRR}_{\bf q} (:R_{q_1}R_{q_2}R_{q_3}:_B+R\rightarrow L) \nonumber
\\ &+&
 \frac{1}{L^3}\sum_{\bf q} \Gamma^{B, RRRL}_{\bf q}\left(:R_{q_1}R_{q_2}R_{q_3}L_{q_4}:_B+R\leftrightarrow L\right)
 \,. \nonumber \\ &&
\label{sec2:HamiltonianBosonFinal}
\end{eqnarray}
The bosonic vertex  $\Gamma^{B, RRRL}_{\bf q}$ has a complicated singular behavior at
small momenta. Specifically, we obtain (see Appendix
\ref{app:HamiltonianBosons})
\begin{equation}
 \Gamma^{B, RRRL}_{\bf q}\approx \tilde{\Gamma}^{B, RRRL}_{\bf q}+l^2
\tilde{\tilde{\Gamma}}^{B, RRRL}_{\bf q}\,, \qquad q^2 l^2\ll 1, 
 \label{sec2:GammaB_RRRL1}
\end{equation}
where
\begin{equation}
\tilde{\Gamma}^{B, RRRL}_{\bf q}=\frac{4 \pi ^3 \alpha}{3 {m^{*}}^2 u_0}  \left[1-\frac{3 \alpha }{2}
-\frac{\alpha}{4}  \left(\frac{ q_4}{q_1}+\frac{ q_4}{q_2}+\frac{ q_4}{q_3}\right)\right],\;
\label{sec2:GammaB_RRRL2}
\end{equation}
and
\begin{eqnarray}
   \tilde{\tilde{\Gamma}}^{B, RRRL}_{\bf q}&=&\frac{ 5\pi ^3 \alpha}{ {m^{*}}^2
u_0}  
  \left[\frac{  q_1 q_2 q_3}{ q_4}+\frac{(6-13 \alpha )}{20}   \left(q_1^2+q_2^2+q_3^2\right)
  \right. \nonumber \\
 &-&  \left.\frac{\alpha}{30} q_4 
\left(\frac{q_1^2+q_2^2}{q_3}+\frac{q_1^2+q_3^2}{q_2}+\frac{q_2^2+q_3^2}{q_1}
\right)
\right. \nonumber \\ 
&+& 
\frac{26\alpha-53 \alpha^2 -8}{60 \alpha} q_4^2 \nonumber \\
   &+&  \left. \frac{1-5 \alpha}{30}  q_4^3 
\left(\frac{1}{q_1}+\frac{1}{q_2}+\frac{1}{q_3}\right)
   \right].
   \label{sec2:GammaB_RRRL3}
\end{eqnarray}

%%%%%%%%%%%%%%%%%%%%%%%%%%%%%%%%%%%%%%%%%%%%%%%%%%%%%%%%%
\begin{figure}
\includegraphics[width=210pt]{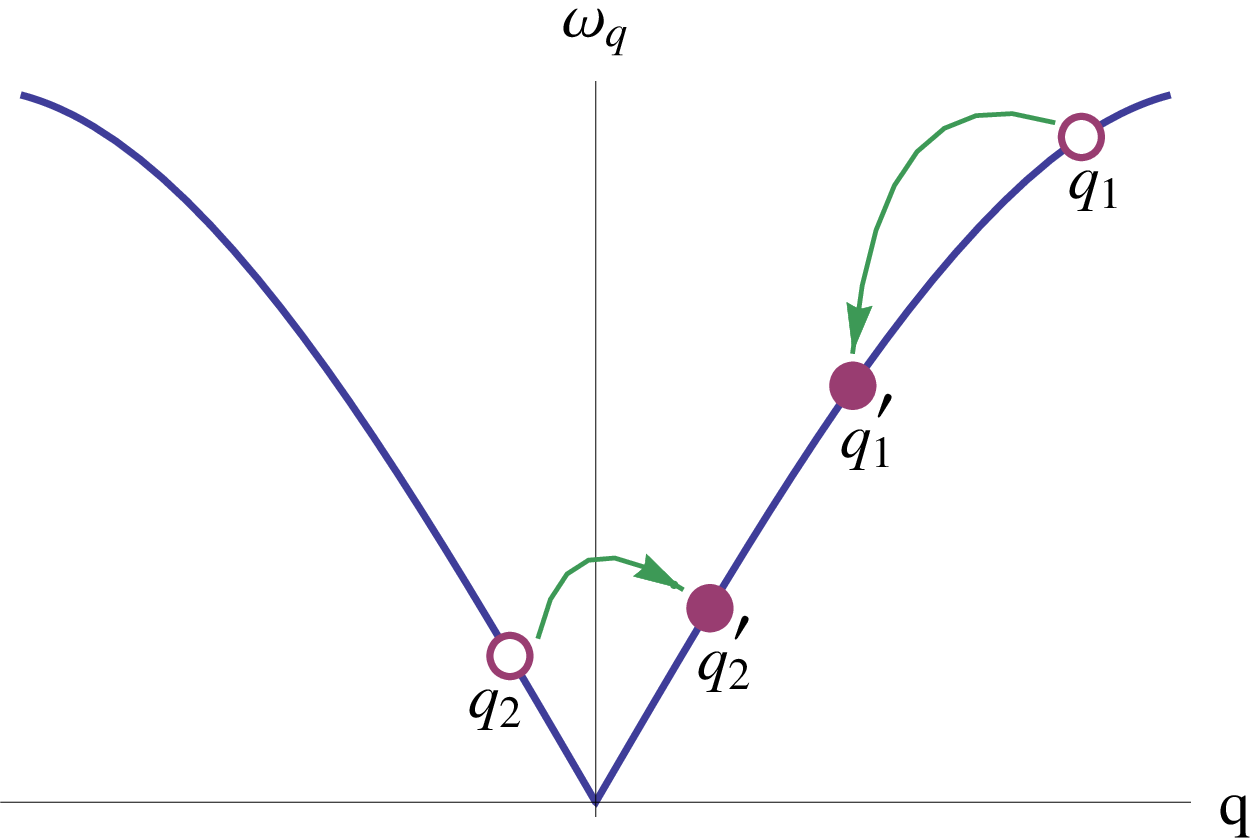}

\vspace{0.3cm}

\includegraphics[width=210pt]{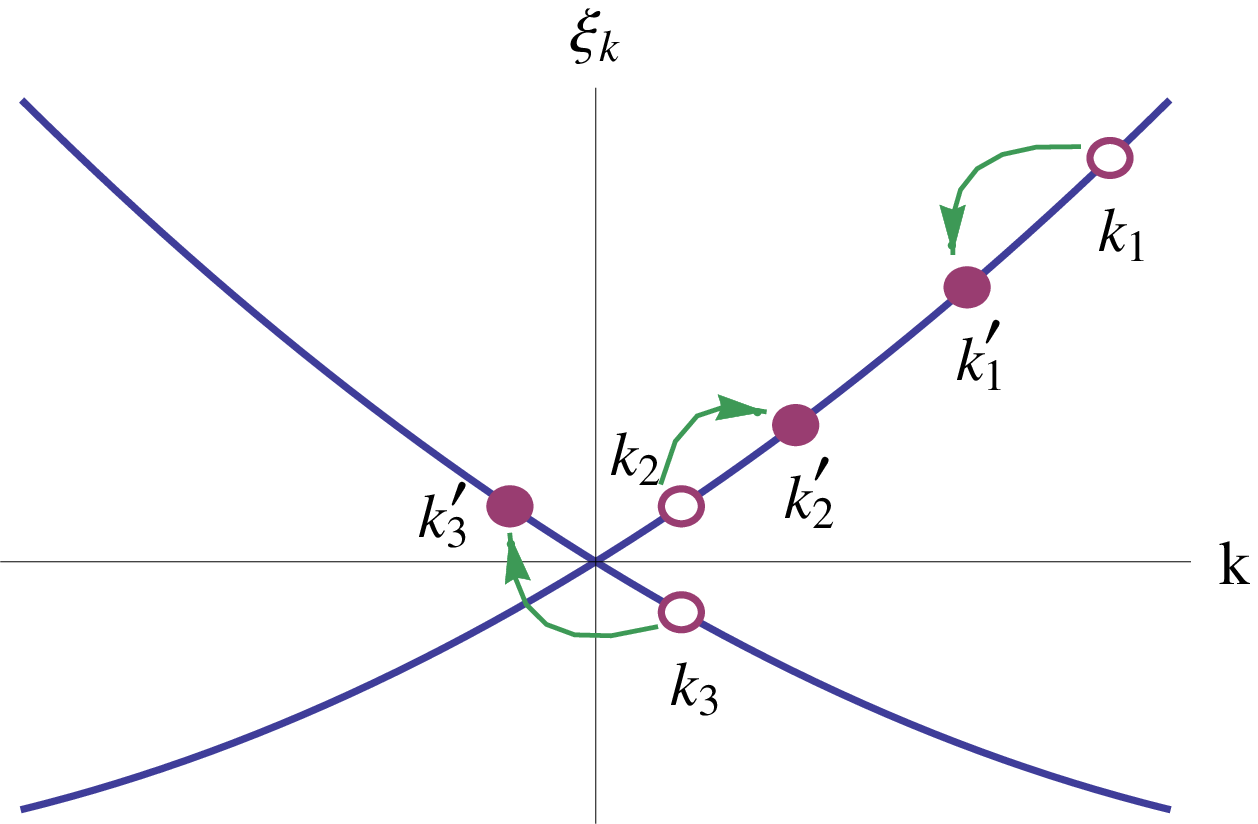}
 \caption{Leading relaxation processes for bosonic (upper panel) and fermionic
(lower panel) excitations.}
 \label{sec2:Fig1}
\end{figure}
%%%%%%%%%%%%%%%%%%%%%%%%%%%%%%%%%%%%%%%%%%%%%%%%%%%%%%%%%%%

In the next sections we will use  the Hamiltonian
(\ref{sec2:HamiltonianBosonFinal})  to study the
the life times of bosons and fermions in 1D interacting system.  The leading
processes contributing to the relaxation of bosonic and fermionic distribution
functions are illustrated in Fig. \ref{sec2:Fig1}. For bosons, this is the 
two-into-two scattering involving a change of the branch for one of the
bosons. 
For fermions, these are three-fermion collisions involving in the initial state
two particle  from one (say, right) branch and one particle from the other (say,
left) branch (see Fig. \ref{sec2:Fig1}). We notice that both these processes
arise already in the first order of perturbation theory in $\Gamma^{B, RRRL}_{\bf q}$
[This is obvious for the case of the bosonic scattering; for fermions this will
become clear in Sec.\ref{sec:Fermions} where we discuss the fermionic form of
the Hamiltonian (\ref{sec2:HamiltonianBosonFinal})]. The momentum $q_4$ in
$\Gamma^{B, RRRL}_{\bf q}$ has the meaning of the momentum transfer between the left
and right chiral sectors in the collision process. The momentum
and energy conservation dictates then the estimates $q_4\sim l^2 p^3$ for
bosonic and $q_4\sim p^2/m$ for fermionic collisions, where $p$ is the typical
momentum of the right particles involved. This observation 
allows one to simplify dramatically the expression for the coupling $\tilde{\tilde{\Gamma}}^{B,
RRRL}_{\bf q}$ by dropping all the terms which do not contain  $q_4$ in the denominator. In
this approximation we get
\begin{equation}
 \Gamma^{B, RRRL}_{\bf q}=\frac{4 \pi ^3 \alpha}{3 {m^{*}}^2 u_0} 
\left[1-\frac{3 \alpha }{2}+\frac{15}{4}l^2\frac{q_1 q_2 q_3}{q_4}\right].
 \label{sec2:GammaB_RRRL_Final}
\end{equation}
The Hamiltonian (\ref{sec2:HamiltonianBosonFinal}) with the coupling $\Gamma^{B,
RRRL}_{\bf q}$ given by (\ref{sec2:GammaB_RRRL_Final}) and its fermionic version
derived in Section \ref{sec:Fermions} constitute our  starting point for the
analysis of the life time of bosonic and fermionic excitations in the generic
Luttinger liquid model.

\section{Lifetime of bosonic excitations}
\label{sec:Bosons}

In this section we exploit the Hamiltonian (\ref{sec2:HamiltonianBosonFinal}) to
study the decay of bosonic excitations in a dispersive Luttinger liquid.

The Fourier components of the densities $R(x)$ and $L(x)$ can be identified with
bosonic creation and annihilation operators via 
\begin{eqnarray}
R_q=\sqrt{\frac{L |q|}{2\pi}}\left(\Theta(q)b_q+\Theta(-q)b^+_{-q}\right),\nonumber\\
L_q=\sqrt{\frac{L |q|}{2\pi}}\left(\Theta(-q)b_q+\Theta(q)b^+_{-q}\right).
\label{sec3:BosonicOperators}
\end{eqnarray}
Let us consider a boson at momentum $Q\gtrsim T/u_0$ injected into the otherwise
equilibrium Lutiinger liquid characterized by  temperature $T$.  
We assume for definiteness that $Q>0$, so that we are dealing with a decay of a
right moving bosonic excitation. 
The dominant collision process limiting the lifetime of the injected boson is a
scattering on  a thermal left-moving boson  at momentum $q<0$ which is
transfered to the right  
branch (see Fig.\ref{sec2:Fig1}). The lifetime of the injected boson is now
given by the out-scattering term of the linearized collision integral 
\begin{eqnarray}
 \frac{1}{\tau_Q(T)}&=&\frac12\int \frac{dq dQ^\prime
dq^\prime}{(2\pi)^3}W_{Qq}^{Q^\prime
q^\prime}N_{B}(\omega_q)\left(N_{B}(\omega_{Q^\prime})+1\right) \nonumber \\
&\times& \left(N_{B}(\omega_{q^\prime})+1\right)
 \Theta(-q) \Theta(Q^\prime)\Theta(q^\prime).
 \label{sec3:tauGeneral}
\end{eqnarray}
Here $N_B(\epsilon)$ is the equilibrium bosonic distribution function.

The transition probability $W_{Qq}^{Q^\prime q^\prime}$ can be expressed via the
corresponding matrix element of the $T$-matrix 
\begin{multline}
 W_{Qq}^{Q^\prime q^\prime}=(2\pi)^2\left|\langle 0|b_{q^\prime} b_{Q^\prime} T
b^+_{Q}b^+_{q} |0\rangle\right|^2 \delta\left(E_{\mathrm i}- E_{\mathrm
f}\right)\\\times
 \delta\left(P_{\mathrm i}-P_{\mathrm f}\right).
\end{multline}
Here the $\delta$-functions express the energy and momentum conservation in the
collision process. 

It is easy to see that the required matrix element $\langle 0|b_{q^\prime}
b_{Q^\prime} T b^+_{Q}b^+_{q} |0\rangle$ arises in the first order of the 
perturbation theory in the coupling $\Gamma^{B, RRRL}_{\bf q}$ and can be  read
off from Eqs.~(\ref{sec2:GammaB_RRRL_Final}) and (\ref{sec3:BosonicOperators}).
The result get substantially simplified since, according to the
conservation laws, the momentum $q$ of the left particle is given by 
\begin{equation}
 q=-\frac{3}{2}QQ^\prime q^\prime l^2+{\mathrm O}(q^5 l^4).
\end{equation}
In view of this relation, the momentum-dependent and momentum-independent terms
in $\Gamma^{B, RRRL}_{\bf q}$, Eq.~(\ref{sec2:GammaB_RRRL_Final}), give
contributions of the same form to the matrix element $\langle 0|b_{q^\prime}
b_{Q^\prime} T b^+_{Q}b^+_{q} |0\rangle$ which finally reads
\begin{equation}
\langle 0|b_{q^\prime} b_{Q^\prime} T b^+_{Q}b^+_{q}
|0\rangle=\frac{\pi\alpha(1+\alpha)}{ m^{*2}u_0}\sqrt{|QQ^\prime q q^\prime|}.
\label{sec3:T}
\end{equation}

Assuming now that the energy of the relaxing boson is much larger than temperature but 
is not too high ($l^2 u_0 Q^3\ll T$), one can replace the thermal factor 
$N_{B}(\omega_q)$ in Eq.~(\ref{sec3:tauGeneral}) by $1/u_0|q|$ and the other two
thermal factors by unity.  Calculating the resulting integral, we find
\begin{equation}
\frac{1}{\tau_{Q}(T)}=
\frac{\pi \alpha^2(1+\alpha)^2}{48 m^{*4}u_0^4}TQ^4.
\label{sec3:tauHot}
\end{equation}
Equation (\ref{sec3:tauHot}) gives the life time of a hot boson in our system 
and leads to the following estimate for a typical relaxation time of thermal
bosons (i.e., those with momenta $Q\sim T/u_0$):
\begin{equation}
 \frac{1}{\tau_{Q\sim T/u_0}(T)}\propto
\frac{\alpha^2(1+\alpha)^2}{m^{*4}u_0^8}T^5\,,
 \label{sec3:tauFinal}
\end{equation}
up to a prefactor of order unity. 

Equation (\ref{sec3:tauFinal}) for the relaxation time of the
bosonic excitations in the dispersive Luttinger liquid constitutes the main
result of this section. The $T^5$ scaling of the
relaxation rate of bosonic excitation has been earlier
obtained\cite{apostolov13,Lin2013} for a strongly interacting Luttinger
liquid which is characterized by Luttinger parameter $K_0\ll 1$ and is close
to the  Wigner crystal. Our derivation is valid for any $K_0$ and thus
represent a generalization of the result of
Ref.~\onlinecite{apostolov13,Lin2013}. We will return to the question of the
range of validity of Eq.~(\ref{sec3:tauFinal}) in Sec.~\ref{sec:Duality}.

\section{Lifetime of fermionic excitations}
\label{sec:Fermions}

\subsection{Refermionization}

The Hilbert space of a 1D chiral bosonic system is isomorphic to the Hilbert
space of a 1D complex chiral fermions (more precisely, to its charge-zero
sector). Correspondingly, the  
Hamiltonian (\ref{sec2:HamiltonianBosonFinal}) can be viewed as a Hamiltonian of fermions $c_{\eta, k}, c_{\eta, k}^+$ introduced via 
\begin{equation}
 R_q=\sum_{k}c^+_{R, k}c_{R, k+q}\,, \qquad L_q=\sum_{k}c^+_{L, k}c_{L, k+q}.
\label{sec4:Fermions}
 \end{equation}
As was discussed in Sec.~\ref{sec:Model}, the fermions are expected to be the
proper excitations at momenta satisfying 
$m^* u_0 l^2 k \ll 1$. We are now going to discuss the life time of these
low-energy fermionic  quasiparticles created by the operators $c_{\eta, k}$.

We need to rephrase the Hamiltonian (\ref{sec2:HamiltonianBosonFinal}) into the
fermionic language. This can be done by substituting  Eq.~(\ref{sec4:Fermions})
into Eq.~(\ref{sec2:HamiltonianBosonFinal}) and performing the normal ordering
of the resulting expression with respect to fermionic modes.
It is obvious from the structure of the bosonic Hamiltonian
(\ref{sec2:HamiltonianBosonFinal}) that in terms of fermions  
\begin{eqnarray}
 H &=& \sum_{k}\xi_{R, k} :c^+_{R, k}c_{R, k}:_F \nonumber \\
 &+& \frac1L\sum_{\bf k}\Gamma^{F, RR}_{\bf k}:c^+_{R, k_1}c^+_{R, k_2}c_{R,
k_2^\prime}c_{R, k_1^\prime}:_F \nonumber\\
  &+& \frac1L\sum_{\bf k}\Gamma^{F, RL}_{\bf k}:c^+_{R, k_1}c^+_{L, k_2}c_{L,
k_2^\prime}c_{R, k_1^\prime}:_F \nonumber \\
  &+& \frac{1}{L^2}\sum_{\bf k}\Gamma^{F, RRR}_{\bf k}:c^+_{R, k_1}c^+_{R,
k_2}c^+_{R, k_3}c_{R, k_3^\prime}c_{R, k_2^\prime}c_{R, k_1^\prime}:_F \nonumber
\\
  &+& \frac{1}{L^2}\sum_{\bf k}\Gamma^{F, RRL}_{\bf k}:c^+_{R, k_1}c^+_{R,
k_2}c^+_{L, k_3}c_{L, k_3^\prime}c_{R, k_2^\prime}c_{R, k_1^\prime}:_F
\nonumber \\
& + & R\longleftrightarrow L +\ldots 
\label{sec4:HamiltonianFermionsFinal}
\end{eqnarray}
Here the dots stand for the four-fermion interaction terms (containing eight
fermionic operators). These terms do not contribute to the three-fermion
collision processes which we aim to discuss in this work and we omit them
altogether. We denote by  ${\bf k}$ in each of vertices $\Gamma^{F, \ldots}_{\bf
k}$ the set of momenta of the fermionic operators involved
(e.g., ${\bf k}=(k_1, k_2, k_3, k_3^\prime, k_2^\prime, k_1^\prime)$  in the
vertex $\Gamma^{F, RRL}_{\bf k}$; note the order of individual momenta in ${\bf
k}$).  

We refer the reader to Appendix \ref{app:HamiltonianFermions} for details of
the derivation of the couplings entering the fermionic Hamiltonian 
(\ref{sec4:HamiltonianFermionsFinal}) and state here the final
results only. First, the  fermionic  single-particle spectrum  $\xi_{\eta, k}$
receives renormalization  from the density-density interaction in
(\ref{sec2:HamiltonianBosonFinal}) and is given by
\begin{equation}
 \xi_{\eta, k}=\frac{k^2}{2m^*}+\eta \int_{0}^k u(k)dk\approx \eta k
u_0+\frac{k^2}{2m^*}-\frac{\eta l^2u_0k^3}{3}.
\label{sec4:xi}
 \end{equation}
Note that the cubic correction to the fermionic spectrum is small compared to
the quadratic one at $k<1/m^* l^2 u_0$ where the fermions are expected to be
proper quasiparticles of the system. 

The two-particle intrabranch interaction in the fermionic Hamiltonian
(\ref{sec4:HamiltonianFermionsFinal}) arises from the intrabranch
density-density interaction in (\ref{sec2:HamiltonianBosonFinal}), 
\begin{equation}
\label{sec4:GammaFRR}
\Gamma^{F, RR}_{\bf k}=\frac{\pi u_0l^2}{2}(k_1-k_2)(k_1^\prime-k_2^\prime).
 \end{equation}
Here we have take into account the expansion of the bosonic velocity $u_q$ at
small momenta. The vertex $\Gamma^{RR}_{\bf k}$ also receives corrections
from the cubic-in-density 
intrabranch bosonic coupling $\Gamma^{B, RRR}_{\bf q}$. These  corrections are
however parametrically small, and we neglect them. 

The interbranch two- and three-particle couplings, which will be most important
for our analysis of the relaxation, both arise from the bosonic vertex
$\Gamma^{B, RRRL}_{\bf q}$, Eqs. (\ref{sec2:HamiltonianBosonFinal}),
(\ref{sec2:GammaB_RRRL_Final}). Remarkably, a singularity at small momentum
transfer $k_3-k_3^\prime$ which might be expected in view of the last 
term  in Eq.~(\ref{sec2:GammaB_RRRL_Final}) does not show up in
$\Gamma^{RL}_{\bf k}$. This vertex  is mostly determined by the first,
momentum-independent  term in (\ref{sec2:GammaB_RRRL_Final}) and is given by
\begin{equation}
 \Gamma^{F, RL}_{\bf k}=\Gamma^{F, LR}_{\bf k}=\frac{\pi \alpha (2-3\alpha)}{2m^{*2}u_0}k_1 k_1^\prime.
 \label{sec4:GammaFRL}
\end{equation}
On the contrary, the three-fermion coupling $\Gamma^{F, RRL}_{\bf k}$ emerges
solely due to the last term in Eq. (\ref{sec2:GammaB_RRRL_Final}),
\begin{eqnarray}
 \Gamma^{F, RRL}_{\bf k} &=& \frac{5 \alpha 
l^2\pi^2(k_1-k_2)(k_1^\prime-k_2^\prime)}{16m^{*2}u_0(k_3-k_3^\prime)} \nonumber
 \\ &\times& \left[(k_1-k_2)^2-(k_1^\prime-k_2^\prime)^2\right].
 \label{sec4:GammaFRRL}
 \end{eqnarray}
and is singular at $k_3-k_3^\prime=0$.
 
 Let us finally comment on the three-particle intrabranch interaction vertex
$\Gamma^{F, RRR}_{\bf k}$. It arises from the cubic intrabranch coupling in the
bosonic Hamiltonian 
 (\ref{sec2:HamiltonianBosonFinal}). By construction, $\Gamma^{F, RRR}_{\bf k}$
is antisymmetric in the three incoming ($k_1$, $k_2$, $k_3$) and the three
outgoing momenta $k_1^\prime$, $k_2^\prime$ and $k_3^\prime$. Since the bosonic
vertex $\Gamma^{B, RRR}_{\bf q}$ is analytic at small momenta, $\Gamma^{F, RRR}_{\bf k}$
should be of the form
 \begin{equation}
  \Gamma^{F, RRR}_{\bf k}\propto \frac{l^6}{m^*}\prod_{i>j}(k_i-k_j)(k_i^\prime-k_j^\prime).
  \label{sec4:GammaFRRR}
 \end{equation}
We thus see that $\Gamma^{F, RRR}_{\bf k}$ is strongly suppressed by a high power of the
momenta. In fact, it is exactly zero withing our approximation for 
the bosonic coupling $\Gamma^{B, RRR}_{\bf q}$, Eq.~(\ref{sec2:GammaB_RRR}). One
needs to retain the sixth-order terms in the expansion of $\Gamma^{B, RRL}_{\bf
q}$ over momentum 
to generate non-zero $\Gamma^{F, RRR}_{\bf k}$. From now on we will largely
ignore $\Gamma^{F,RRR}_{\bf k}$ apart from a short discussion of the
intrabranch fermionic relaxation processes at the end of the
Sec.~\ref{sec:FermionicScatteringRate}.

With the fermionic Hamiltonian (\ref{sec4:HamiltonianFermionsFinal}) and the
couplings (\ref{sec4:GammaFRR}), (\ref{sec4:GammaFRL}), (\ref{sec4:GammaFRRL})
at hand, we are now in a position to 
study the life time of the fermionic excitations in our problem. Below we employ
the perturbation theory in the fermionic  interaction terms  $\Gamma^{F,
\eta\eta'}_{\bf k}$ and  $\Gamma^{F, RRL}_{\bf k}$ to evaluate the corresponding
scattering rate. 

\subsection{Fermionic scattering rate}
\label{sec:FermionicScatteringRate}

Here we calculate the life time of fermionic excitations in a dispersive
Luttinger liquid. The relaxation of the fermionic distribution function  is
governed by the three-particle collisions. At zero temperature the intrabranch
collision processes (with all three particles in the initial and final states
residing on the right branch) are ruled out by the energy and momentum
conservation. At finite temperature, the situation we consider in this work,
both intrabranch collisions and the scattering events involving the creation of
a particle-hole pair in the left branch (see Fig. \ref{sec2:Fig1}) contribute to
the relaxation of a right-moving fermion injected into the system.  However, the
intrabranch collision rate 
$1/\tau^{RRR}(T)$ turns out to be proportional to a very high power of
temperature ($T^{14}$) and is small compared to the interbranch collision
rate $1/\tau^{RRL}(T)$ in the whole ``fermionic'' part of the parameter space. 
We will discuss this point in more detail at the end of this section and
concentrate now on the contribution of interbranch collisions to the decay of
fermionic quasiparticles. 

The decay rate $1/\tau_{k_1}^{RRL}(T)$ of a fermion with momentum $k_1\gtrsim T/u_0$  is given by the out-scattering term of the linearized three-particle collision integral,
\begin{eqnarray}
 1/\tau^{RRL}_{k_1}(T) &=&\frac12 \int \left(d {\bf k}\right)W_{k_1, k_2,
k_3}^{k_1^\prime, k_2^\prime, k_3^\prime}N_F(\xi_{R, k_2})
 N_F(\xi_{L, k_3}) \nonumber \\
 &\times& \left(1-N_F(\xi_{R, k_1^\prime})\right)\left(1-N_F(\xi_{R,
k_2^\prime})\right) \nonumber \\ &\times& \left(1-N_F(\xi_{L,
k_3^\prime})\right).
\label{sec4:tauRRL_General}
 \end{eqnarray}
Here $(d{\bf k})=d k_2 dk_3 dk_1^\prime dk_2^\prime dk_3^\prime/(2\pi)^5$ and $N_F(\epsilon)$ stands for  the Fermi-Dirac distribution at temperature $T$. 
The transition probability $W_{k_1, k_2, k_3}^{k_1^\prime, k_2^\prime,
k_3^\prime}$  entering Eq. (\ref{sec4:tauRRL_General}) is given by the modulus
squared of the appropriate entry of the $T$-matrix, 
\begin{eqnarray}
 W_{k_1, k_2, k_3}^{k_1^\prime, k_2^\prime, k_3^\prime} &=& (2\pi)^2
\left|\langle 1, 2, 3|T| 1^\prime, 2^\prime,
3^\prime\rangle\right|^2 \nonumber \\
&\times& \delta\left(E_{\mathrm i}-E_{\mathrm
f}\right) \delta\left(P_{\mathrm i}-P_{\mathrm f}\right)
 \,.
 \label{sec4:W}
\end{eqnarray}
Here the $\delta$-functions express the conservation of energy and momentum in the collision process and 
\begin{equation}
 |1, 2, 3\rangle=c^+_{R, k_1}c^+_{R, k_2}c^+_{L, k_3}|0\rangle.
\end{equation}

Examination of the fermionic Hamiltonian (\ref{sec4:HamiltonianFermionsFinal})
shows that to the leading order in $1/m$ there are two contributions 
to the matrix element $\langle 1, 2, 3|T| 1^\prime, 2^\prime, 3^\prime\rangle$.
The first one stems from the three-fermion coupling $\Gamma^{F, RRL}_{k}$ in the
first order of perturbation theory and is given by 
\begin{equation}
 \langle 1, 2, 3|T| 1^\prime, 2^\prime, 3^\prime\rangle_1=4\Gamma^{F, RRL}_{k_1^\prime, k_2^\prime, k_3^\prime, k_3, k_1, k_2},
 \label{sec4:T1General}
\end{equation}
where the vertex $\Gamma^{F, RRL}_{k}$ is given by (\ref{sec4:GammaFRRL}).
The second contribution arises in the second order of the perturbation theory in
the two-fermion couplings $\Gamma^{RR}_{\bf k}$ and $\Gamma^{RL}_{\bf k}$, 
\begin{multline}
 \langle 1, 2, 3|T| 1^\prime, 2^\prime, 3^\prime\rangle_2\\=\frac{8\Gamma^{F, RR}_{k_1^\prime, k_2^\prime, k_2, q}
 \left(\Gamma^{F, RL}_{q, k_3^\prime, k_3, k_1}+\Gamma^{F, LR}_{q, k_3^\prime, k_3, k_1}\right)}{\xi_{L, k_3}-\xi_{L, k_3^\prime}+\xi_{R, k_1}-\xi_{R, q}}\\
 +\frac{8\Gamma^{F, RR}_{k_1^\prime, q, k_2, k_1}
 \left(\Gamma^{F, RL}_{k_2^\prime, k_3^\prime, k_3, q}+\Gamma^{F, LR}_{k_2^\prime, k_3^\prime, k_3, q}\right)}{\xi_{L, k_3^\prime}-\xi_{L, k_3}+\xi_{R, k_2^\prime}-\xi_{R, q}}.
\label{sec4:T2General}
 \end{multline}
 In Eq.~(\ref{sec4:T2General}) we implicitly assume the anti-symmetrization of
the right-hand side with respect to permutations of $k_1$ and $k_2$ as well as
$k_1^\prime$ and $k_2^\prime$. The momentum $q$ of the intermediate  virtual
state in Eq.~(\ref{sec4:T2General}) is fixed by the momentum conservation in
the vertices $\Gamma^{F, RL}_{\bf k}$ implying that  
$q=k_1+k_3-k_3^\prime$ for the first term and $q=k_2^\prime+k_3^\prime-k_3$ for the second one. 

Let us now consider the energy denominators in (\ref{sec4:T2General}) in more
detail. Using the explicit form (\ref{sec4:xi}) of the single-particle
dispersion relations, one finds for the first energy denominator
\begin{multline}
 \xi_{L, k_3}-\xi_{L, k_3^\prime}+\xi_{R, k_1}-\xi_{R, q}\\=-2 u_0
(k_3-k_3^\prime)[1+O(k/p_F)+O(k^2 l^2 )],
\end{multline}
with $k$ being the characteristic value of the momenta. The second denominator in (\ref{sec4:T2General}) has exactly the same structure. 
Working to the leading order in $k/p_F$ , $k^2 l^2$ and using explicit
expressions for the vertices $\Gamma^{F, RR}_{\bf k}$, $\Gamma^{F, RL}_{\bf k}$ derived earlier,
we get (after the proper anti-symmetrization over momenta)
\begin{multline}
 \langle 1, 2, 3|T| 1^\prime, 2^\prime, 3^\prime\rangle_2\\
 =\frac{\pi^2\alpha(2-3\alpha)l^2(k_1-k_2)(k_1^\prime-k_2^\prime)}{2m^{*2}u_0(k_3-k_3^\prime)}
 \\
\times \left(k_1^2 +k_2^2+4k_1k_2-k_1^{\prime2}-k_2^{\prime 2}-4k_1^\prime
k_2^\prime\right).
\label{sec4:T2GeneralSimplified}
 \end{multline}

Let us assume from now on that the the temperature of the system is low, $ m^*
l^2 T\ll 1$, i.e, we are in the situation when the fermions are expected to be
the proper quasiparticles for the description of the system. Under this
condition we can neglect the cubic term in the fermionic  dispersion relation
(\ref{sec4:xi}).
The matrix elements (\ref{sec4:T1General}) and (\ref{sec4:T2GeneralSimplified})
can be further simplified if one takes into account the energy  conservation in
the collision process. 
First, we note that the energy and momentum conservation requires that at zero
momentum transfer $k_3-k_3^\prime=0$ the right-moving particles can only
preserve or exchange their momenta. Thus, the singularity in the matrix elements
 (\ref{sec4:T1General}) and (\ref{sec4:T2GeneralSimplified}) is canceled  if we
consider them on the mass shell. Second, on the mass shell the momentum transfer
can be estimated as $k_3-k_3^\prime\sim k^2/m^* u_0$, where $k$ is the
characteristic momentum of the colliding particles. Consequently, 
we can estimate the matrix elements (\ref{sec4:T1General}) and (\ref{sec4:T2GeneralSimplified}) on the mass shell as
\begin{equation}
 \langle 1, 2, 3|T| 1^\prime, 2^\prime, 3^\prime\rangle_{1 (2)}\propto
\frac{l^2(k_1-k_2)(k_1^\prime-k_2^\prime)}{m^*}.
\end{equation}
The accurate calculation presented in Appendix \ref{app:MassShellAmplitudes}
confirms this estimate and yields\cite{Remark_T}
\begin{equation}
 \langle 1, 2, 3|T| 1^\prime, 2^\prime, 3^\prime\rangle=
 \frac{6\pi^2\alpha(1+\alpha)l^2(k_1-k_2)(k_1^\prime-k_2^\prime)}{m^{*}}.
 \label{sec4:T}
\end{equation}

Assuming now that the energy of the relaxing particle $u_0 k_1$  is not too high ($u_0 k_1\ll \sqrt{T u_0^2 m^*}$) one can linearize the fermionic spectra in the energy conserving $\delta$-function in Eq.(\ref{sec4:W}). The integration over $k_3$ and $k_3^\prime$ is Eq.(\ref{sec4:tauRRL_General}) is then straightforward and leads to
\begin{multline}
 1/\tau^{RRL}_{k_1}(T)=\frac{T}{8\pi u_0^2}\int
 \frac{dk_2 d k_1^\prime dk_2^\prime}{(2\pi)^3}T_{k_1, k_2}^{k_1^\prime, k_2^\prime}
 N_F(\xi_{R, k_2^\prime})
 \\
 \times
   \left(1-N_F(\xi_{R, k_1^\prime})\right)\left(1-N_F(\xi_{R, k_2^\prime})\right)
   2\pi \delta(k_1+k_2-k_1^\prime-k_2^\prime)
\label{sec4:tauRRL_Simplified}
 \end{multline}
At $k_1\gg T/u_0$ we can approximate the Fermi distributions entering
(\ref{sec4:tauRRL_Simplified}) by the zero-temperature ones, which leads to the
following result for the relaxation rate of a hot fermion in our system: 
\begin{equation}
 1/\tau^{RRL}_{k_1}(T)=\frac{11\pi\alpha^2(1+\alpha)^2l^4 Tk_1^6}{80m^* u_0^2}.
\end{equation}
The corresponding estimate for the lifetime of the thermal quasiparticles with $k_1\sim T/u_0$ constitutes the central result of this section
\begin{equation}
 1/\tau^{RRL}_{k_1\sim T/u_0}(T)\propto \frac{\alpha^2(1+\alpha)^2 l^4
T^7}{m^{*2}u_0^8}.
\label{sec4:tauRRL}
 \end{equation}
 
Equation (\ref{sec4:tauRRL}) establishes the relaxation rate of the fermionic
quasiparticles in a general dispersive Luttinger liquid. 
Its scaling with temperature coincides with the one  obtained previously  for
weakly interacting fermions within a perturbation theory \cite{imambekov11} and
for fermionic quasiparticles of a Luttinger liquid with a short-range
interaction \cite{MatveevFurusaki2013}. The $T^7$ scaling 
in Eq.~(\ref{sec4:tauRRL}) can be traced back to a product of $T^4$ factor
arising due to the quadratic scaling of the matrix element (\ref{sec4:T}) with
the momentum and a $T^3$ factor steming from the phase volume. 

We return now to the scattering rate $1/\tau^{RRR}_{k_1}(T)$ induced by
intrabranch three-particle collisions. The corresponding amplitude
$\langle 123| T|1^\prime2^\prime3^\prime\rangle$ (with all the particle
belonging now to the right branch) arises in the first order of the perturbation
theory over the vertex $\Gamma^{F, RRR}_{\bf k}$ as well as in the second order
in the intrabranch two-particle interaction  $\Gamma^{F, RR}_{\bf k}$. 
The matrix element induced by $\Gamma^{F, RRR}_{\bf k}$  is given by
[see Eq.~(\ref{sec4:GammaFRRR})]
\begin{multline}
  \langle 123| T|1^\prime2^\prime3^\prime\rangle_1\propto \frac{l^6}{m^*}\prod_{i>j}(k_i-k_j)(k_i^\prime-k_j^\prime).
  \label{sec4:T1RRR}
\end{multline}
A careful examination of the second order of the perturbation theory in
$\Gamma^{F, RR}_{\bf k}$ shows that, despite the presence of energy
denominators, the corresponding matrix element 
$ \langle 123| T|1^\prime2^\prime3^\prime\rangle_2$ is non-singular at the mass
shell and has the same momentum dependence (dictated by indistinguishability of
the particles) as Eq.~(\ref{sec4:T1RRR}),
\begin{equation}
 \langle 123| T|1^\prime2^\prime3^\prime\rangle_2\propto m^* u_0^2 l^8
\prod_{i>j}(k_i-k_j)(k_i^\prime-k_j^\prime).
\label{sec4:T2RRR}
 \end{equation}
To obtain Eq. (\ref{sec4:T2RRR}) one has to go beyond the approximation
(\ref{sec2:uq}) for the momentum-dependent bosonic velocity and the
corresponding approximation 
(\ref{sec4:GammaFRR}) for the intrabranch two-particle interaction
$\Gamma^{RR}_{\bf k}$. Specifically, one has to retain the $O(k^4)$ terms for
both vertices $\Gamma^{RR}_{k}$
involved. This is the reason for the appearance of the factor $l^8$ in
(\ref{sec4:T2RRR}). The factor of mass $m^*$ in Eq. (\ref{sec4:T2RRR}) comes
from the energy denominator of the second-order perturbation theory and reflects
its degenerate nature for dispersionless fermions. 

Comparing Eqs. (\ref{sec4:T1RRR}) and (\ref{sec4:T2RRR}), we observe that the
second contribution dominates due to an additional factor $ (m^* u_0 l)^2
\gtrsim 1$, so that the matrix element for the 
intrabranch triple collisions is given by Eq. (\ref{sec4:T2RRR}).
The evaluation of the intrabranch transition rate is now a matter of power
counting resulting in
\begin{equation}
 1/\tau^{RRR}_{k_1\sim T/u_0}(T)\propto \frac{m^{*3} l^{16} T^{14}}{u_0^{10}}.
 \label{sec4:tauRRR}
\end{equation}
The second power of mass in Eq. (\ref{sec4:tauRRR}) comes from the matrix
element (\ref{sec4:T2RRR}) and an additional factor $m^*$ arises from the
$\delta$-function expressing the energy conservation due to the fact that energy
and momentum conservation coincide for particles with the linear spectrum. 
Comparing (\ref{sec4:tauRRR}) to (\ref{sec4:tauRRL}), we see that the
interbranch collision processes dominate in the entire range of temperatures
$m^*l^2 T/u_0\ll 1$ where the above fermionic analysis is justified.

\section{Fermi-Bose weak-strong-coupling duality}
\label{sec:Duality}

In the previous sections we have presented a detailed analysis  of 
relaxation times of bosonic and fermionic excitations in a dispersive
Luttinger liquid. For this purpose, we have carried out a perturbative
treatment of the Hamiltonian (\ref{sec2:HamiltonianBosonFinal}) and of its
fermionized version (\ref{sec4:HamiltonianFermionsFinal}), respectively. 
Comparing now the two calculations above, we observe that the
fermionic and bosonic relaxations are closely related: they both
originate from the same interaction term 
$\Gamma^{RRRL}_{\bf q}$ in the Hamiltonian (\ref{sec2:HamiltonianBosonFinal})
and are both dominated by the processes with small momentum transfer between the
right and left chiral branches.  An important difference between the bosonic and
fermionic scattering processes is the scaling of this momentum transfer with
the typical momentum of the right particles, which is cubic for bosons and
quadratic for fermions.  

Comparison of the bosonic and fermionic  relaxation times,
Eqs.~(\ref{sec3:tauFinal}) and (\ref{sec4:tauRRL}), reveals the
dimensionless parameter $\lambda=m^* l^2 T$ anticipated in
Sec.~\ref{sec:Model}. In agreement with the qualitative discussion in
Sec.~\ref{sec:Model}, at low temperatures, $\lambda\ll 1$, the
result (\ref{sec4:tauRRL}) for the decay rate of fermionic excitations
in a dispersive Luttinger liquid is much smaller than 
Eq.~(\ref{sec3:tauFinal}) resulting from a bosonic perturbative treatment of the
Hamiltionian (\ref{sec2:HamiltonianBosonFinal}). Thus, fermions are proper
excitations in this regime. The situation is reverse at high temperatures where
$\lambda\gg 1$.  

To support this subdivision of the parameter space in ``fermionic'' and
``bosonic'' domains, let us analyze the perturbation theories used in the
previous sections to evaluate the bosonic and fermionic lifetimes. We consider
first the fermionic formalism. As follows from Eq.~(\ref{sec4:xi}), a finite
interaction radius $l_{\rm int}$ induces a cubic correction to the spectrum
of the fermionic quasiparticles. This correction is small in comparison
to the original curvature $1/m^*$ at momenta
$k\ll 1/m^* u_0 l^2$, yielding the condition $\lambda\ll  1$ for the
fermionic perturbation theory. Furthermore, an estimate for the scaling of
higher-order diagrams (with 4, 5, \ldots fermions involved) confirms that the
perturbation theory is controlled by the parameter $\lambda\ll  1$.
Conversely, a finite fermionic mass broadens the
support of the dynamical structure factor in the frequency-momentum plane
$(\omega, q)$ by an amount of the order of $\delta\omega\sim q^2/m^*$. This
broadening exceeds the nonlinear bending of the bosonic single-particle spectrum
$u_0l^2 q^3$ at momenta $q\lesssim 1/m^* u_0l^2$ and makes the perturbative
treatment of the bosonic Hamiltonian inadequate for $\lambda\lesssim 1$. In
other words, a small parameter controlling the bosonic perturbation theory is
$\lambda^{-1} \ll 1$. To summarize, the fermionic and bosonic 
descriptions of a dispersive Luttinger liquid are characterized by the
coupling constants $\lambda$ and $\lambda^{-1}$, respectively, thus showing a
remarkable weak-strong-coupling duality.

\begin{figure}
\includegraphics[width=220pt]{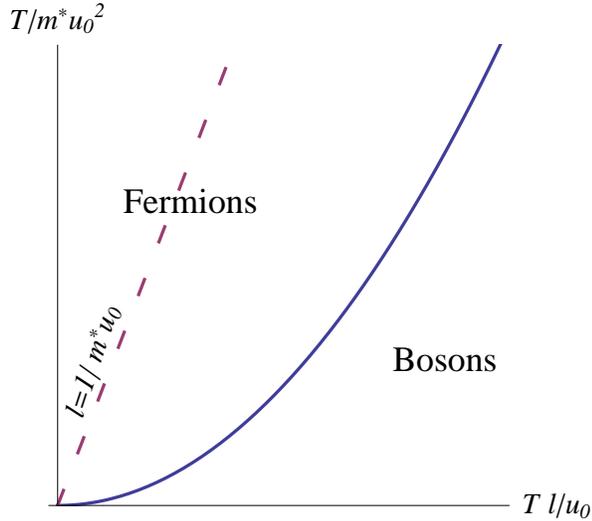}
\caption{\small ``Phase diagram'' of a dispersive Luttinger liquid in the
parameter plane $(x=T l/u_0,\ y=T/m^*u_0^2)$.
Relaxation rates in the bosonic and fermionic parts of
the phase diagram are given by Eqs.~(\ref{sec3:tauFinal}) and
(\ref{sec4:tauRRL}), respectively.
The solid line $y=x^2$
indicates a crossover between the fermionic and bosonic regimes. The dashed
line corresponds to the minimal physically sensible interaction length $l \sim
1/m^*u_0$. } 
\label{Fig:Diagram}
\end{figure}

The ``phase diagram'' of a dispersive Luttinger liquid exhibiting a crossover
between the bosonic and fermionic  regimes is shown in Fig. \ref{Fig:Diagram}
in the coordinates $(T l/u_0,\ T/m^*u_0^2)$. 

%%%%%%%%%%%%%%%%%%%%%%%%%%%%%%%%%%%%%%%%%%%%%%%%%%%%%%%%%%%%%%%%%%%%%%%%%%%%
%%%%%%%%%%%%%%%%%%%%%%%%%%%%%%%%%%%%%%%%%%%%%%%%%%%%%%%%%%%%%%%%%%%%%%%%%%%

\section{Summary and outlook}
\label{sec:Summary}

To summarize, we have explored the life time of excitations in a dispersive
Luttinger liquid in the whole range of parameters. 
We employed   bosonization approach  supplemented by a sequence of unitary
transformations to a quasiparticle representations which allowed us to eliminate
many of interaction-induced contributions from the Hamiltonian.  The resulting 
bosonic Hamiltonian is given by Eq.~(\ref{sec2:HamiltonianBosonFinal}) and its
refermionized version by Eq.~(\ref{sec4:HamiltonianFermionsFinal}).

We have performed both bosonic and fermionic analysis of the relaxation rate in
this formalism.  The central results of this work, Eqs. (\ref{sec3:tauFinal})
and (\ref{sec4:tauRRL}), reveal the Bose-Fermi weak-strong coupling duality
controlled by the parameter $\lambda = m^*l^2T$ and allow us to establish the
``Bose-Fermi phase diagram''  of a generic dispersive Luttinger liquid presented
in Fig.~\ref{Fig:Diagram}. 

Remarkably, the parameter $\lambda$ controlling the Bose-Fermi crossover in the
relaxation mechanisms studied in this work is closely
related to the parameter $\lambda_\rho=u_0 m^*l^2\Delta \rho$ which was shown
recently\cite{PGM2013,protopopov13} to govern the character of the collisionless
evolution 
of a density perturbation with an amplitude $\Delta \rho$ in a dispersive
Luttinger liquid.  Specifically, it was found in Ref.~\onlinecite{PGM2013} that
for $\lambda_\rho \ll 1$ (``fermionic'' regime) the corresponding collisionless
kinetic equation predicts a formation of the population inversion in the
distribution function of fermions while for $\lambda_\rho\gg 1$ (``bosonic''
regime) no such phenomenon occurs and the density evolution follows closely
the predictions of a hydrodynamic theory.  Comparing the expressions for
$\lambda$ and $\lambda_\rho$, we observe that they are identical, up to a
replacement of the characteristic energy scale $T$ by $u_0 \Delta\rho$.
In this context, our present findings  open up the possibility to incorporate
the relaxation processes into the description of the pulse propagation in 
dispersive Luttinger liquids.  In the ``fermionic'' regime the fermionic
collisions studied in this work can be directly included into kinetic equation
of  Ref.~\onlinecite{PGM2013}. On the other hand, a proper account for the
relaxation processes in the ``bosonic'' regime of the pulse propagation requires
a formulation of a bosonic version of the kinetic equation, which remains a
prospect for future research. 

Another direction for future work is the investigation of a broader
class of interaction potentials within our formalism. In particular, it would
be interesting to study the Fermi-Bose duality in relaxation of excitations
in the case of power-law ($1/r^\alpha$) interactions. Especially important, in
view of applications to charged fermions, is the case of Coulomb ($1/r$)
interaction screened (e.g., due to a remote gate) at a large distance $d \gg
p_F^{-1}$. We also envision an extension of our approach to the situation when
the initial interacting particles are bosons, which is relevant in the context
of the physics of cold atoms in one-dimensional traps.

\begin{acknowledgments}

We acknowledge useful discussions with I.V. Gornyi, A. Levchenko, and D.G.
Polyakov, and financial support by Israeli Science Foundation, by German-Israeli
Foundation, and by DFG Priority Program 1666. 

\end{acknowledgments}

\vskip0.5cm

\appendix 

\section{Unitary transformation $U_3$ and bosonic vortexes}
\label{app:HamiltonianBosons}

In this Appendix we derive explicit expressions for the bosonic vertices
$\Gamma^{B, RRRR}_{\bf q}$, $\Gamma^{B, RRRL}_{\bf q}$ and $\Gamma^{B, RRLL}_{\bf q}$.
Our starting point is the Hamiltonian after the unitary rotation $U_2$, see
Eq.~(\ref{sec2:HamiltonianBoson1}). In terms of the new density operators 
$\tilde{\rho}_\eta$ the Hamiltonian reads
\begin{eqnarray}
 H&=&(\pi/L)\sum_{q}u_q:\left(
\tilde{\rho}_{R, q}\tilde{\rho}_{R, -q}+
\trho_{L, q}\trho_{L, -q}\right):_B \nonumber \\
&+& \frac{1}{L^2}\sum_{{\bf q}}
 \left[\Gamma^{B, RRR}_{\bf q} :(\trho_{R, q_1}\trho_{R, q_2}\trho_{R,
q_3}+R\rightarrow L):_B \right. \nonumber \\ &+& \left.
      \Gamma^{B, RRL}_{\bf q} :(\trho_{R, q_1}\trho_{R, q_2}\trho_{L, q_3}+R\leftrightarrow L):_B
\right] \nonumber \\
&\equiv& H^{(2)}+H^{(3)}_{D} + H^{(3)}_{O}
\,.
\label{app1:HamiltonianBoson1}
\end{eqnarray}
Here we have splitted the Hamiltonian into the quadratic part $H^{(2)}$, the
diagonal-in-chiralities cubic part $H^{(3)}_{D}$ and the chirality-mixing cubic
part $H^{(3)}_{O}$.

Expressing now the densities $\trho_{R(L)}$ via
Eq.~(\ref{sec2:DensityExpansion}), we find
\begin{eqnarray}
 H &=& H^{(2)}+H^{(3)}_D+H^{(3)}_{O}-\left[\Omega_3,
H^{(2)}\right] \nonumber \\ &-&\left[\Omega_3,
H^{(3)}_{D}+H^{(3)}_{O}\right]+\frac12 \left[\Omega_3, \left[\Omega_3,
H^{(2)}\right]\right] \nonumber \\ &+&
 {\mathrm O}(\rho^5)\,.
 \label{app1:HamiltonianBoson21}
\end{eqnarray}
The operators $H^{(2)}$, $H^{(3)}_{D(O)}$ in the right hand side of
Eq.(\ref{app1:HamiltonianBoson21}) are obtained from that of
Eq.(\ref{app1:HamiltonianBoson1}) by a simple replacement
$\trho_{R(L)}\rightarrow R(L)$.

The decoupling of the chiral sectors in the third order requires that
\begin{equation}
 \left[\Omega_3, H^{(2)}\right]=H^{(3)}_{O}\,,
 \label{app1:f}
\end{equation}
which is equivalent to Eq.~(\ref{sec2:f}). Using Eq. (\ref{app1:f}), one can
bring Eq.~(\ref{app1:HamiltonianBoson21}) to a simpler form
\begin{equation}
 H=H^{(2)}+H^{(3)}_D-\left[\Omega_3, H^{(3)}_{D}+\frac12 H^{(3)}_{O}\right]+
 {\mathrm O}(\rho^5)\,.
 \label{app1:HamiltonianBoson22}
\end{equation}
Computing the commutator, we obtain the following result for the fourth-order
correction to the Hamiltonian:
\begin{widetext}
\begin{eqnarray}
 H^{(4)}&=&\frac1{2\pi L^3}\sum_{{\bf q}, p}
 \left[-6 p\Gamma^{B, RRR}_{-p, q_1, q_2}f_{p, q_3, q_4}+p\Gamma^{B, RRL}_{-p, q_4, q_3}f_{q_1, q_2, p}
 -p\Gamma^{B, RRL}_{q_1, q_2, -p}f_{p, q_4, q_3}
 \right]\left[\left(R_{q_1}R_{q_2}R_{q_3}\right)_s L_{q_4}+R\leftrightarrow
L\right] \nonumber \\
&+& \frac{1}{2\pi L^3}\sum_{{\bf q}, p}\left[3p\Gamma^{B, RRR}_{-p, q_1,
q_2}f_{q_3, q_4, p}+3p\Gamma^{B, RRR}_{-p, q_3, q_4}f_{q_1, q_2, p}
 -2 p\Gamma^{B, RRL}_{-p, q_1, q_3}f_{p, q_2, q_4}-2 p\Gamma^{B, RRL}_{-p, q_3, q_1}f_{p, q_4, q_2}
 \right]
 \left(R_{q_1}R_{q_2}\right)_s\left(L_{q_3}L_{q_4}\right)_s \nonumber \\
&+& \frac{1}{4\pi L^3}\sum_{{\bf q}, p}p \Gamma^{B,RRL }_{q_1, q_2, -p}f_{q_3,
q_4, p}\left[\left(R_{q_1}R_{q_2}R_{q_3}R_{q_4}\right)_s
+R\rightarrow L
\right].
\label{app:HamiltonianBoson2}
\end{eqnarray}
\end{widetext}
Here the subscript $s$ in the expressions of the type $\left(\ldots\right)_s$
stands for a symmetrization of the expression inside brackets with respect to
the momenta $ q_i$. 

Comparing Eq.~(\ref{app:HamiltonianBoson2}) to
Eq.~(\ref{sec2:HamiltonianBoson2}), one can read off explicit expressions
for fourth-order bosonic vertices [$\Gamma^{B, RRRR}_{\bf q}$ etc.] in terms of
$\Gamma^{B, RRR}_{\bf q}$ and $\Gamma^{B, RRL}_{\bf q}$. Exploiting the
expansion of third-order vertices at small momenta, we get
\begin{equation}
 \Gamma^{B, \mu\nu\rho\kappa}_{\bf q}\approx \tilde{\Gamma}^{B,
\mu\nu\rho\kappa}_{\bf q}+l^2 \tilde{\tilde{\Gamma}}^{B, \mu\nu\rho\kappa}_{\bf
q}\,, \qquad q^2 l^2\ll 1\,,
\end{equation}
with 
\begin{eqnarray}
\tilde{\Gamma}^{B, RRRR}_{\bf q} &=& -\frac{\pi^3\alpha^2}{2 u_0 {m^*}^2},
\qquad \\
\tilde{\Gamma}^{B, RRRL}_{\bf q} &=& \frac{4 \pi ^3 \alpha}{3 {m^{*}}^2 u_0} 
\left[1-\frac{3 \alpha }{2}
-\frac{\alpha}{4}  \left(\frac{ q_4}{q_1}+\frac{ q_4}{q_2}+\frac{ q_4}{q_3}\right)\right], \nonumber\\
\end{eqnarray}
and
\begin{eqnarray}
   \tilde{\tilde{\Gamma}}^{B, RRRL}_{\bf q} &=& \frac{ 5\pi ^3 \alpha}{
{m^{*}}^2 u_0} \nonumber \\  
&\times&  \left[\frac{  q_1 q_2 q_3}{ q_4}+\frac{(6-13 \alpha )}{20}  
\left(q_1^2+q_2^2+q_3^2\right)
  \right. \nonumber \\ &-& \left. \frac{\alpha}{30} q_4 
\left(\frac{q_1^2+q_2^2}{q_3}+\frac{q_1^2+q_3^2}{q_2}+\frac{q_2^2+q_3^2}{q_1}
\right)
\right. \nonumber \\ &+& 
\frac{26\alpha-53 \alpha^2 -8}{60 \alpha} q_4^2 \nonumber \\
   &+& \left. \frac{1-5 \alpha}{30}  q_4^3 
\left(\frac{1}{q_1}+\frac{1}{q_2}+\frac{1}{q_3}\right)
   \right]. 
\end{eqnarray}
For completeness we present also the amplitude $\Gamma^{B, RRLL}_{\bf q}$
although we do not need it in the main text:
 \begin{eqnarray}
   \tilde{\Gamma}^{B, RRLL}_{\bf q} &=&
 \frac{ \pi ^3 \alpha}{{m^*}^2 u_0}  \left[1
+\frac{\alpha\left(q_1-q_2\right)^2}{2q_1 q_2}\right]  +R\leftrightarrow L,
\nonumber \\ && \\
\tilde{\tilde{\Gamma}}^{B,RRLL}_{\bf q} &=&
 \frac{ \pi ^3 \alpha}{{8m^*}^2 u_0 L^3}  \left[
(16-47 \alpha )   \left(q_1^2+q_2^2\right)\right. \nonumber \\
&-& 
\frac{\left(4-3 \alpha ^2\right)}{\alpha }  (q_1+q_2)^2 \nonumber \\
&+& 
17\alpha   \left(q_1^2+q_2^2\right)
   \left(\frac{q_3}{q_4}+\frac{q_4}{q_3}\right) \nonumber \\ &-& \left.
     (5 \alpha +2)   \left(\frac{q_1^3}{q_2}+\frac{q_2^3}{q_1}\right)
   \right] \nonumber \\ &+& R\leftrightarrow L\,.
 \end{eqnarray}

\section{Fermionic form of the Hamiltonian}
\label{app:HamiltonianFermions}

In this Appendix we present a detailed derivation of the fermionized form of the
bosonic Hamiltonian (\ref{sec2:HamiltonianBosonFinal}). 
It is obvious from the structure of the Hamiltonian
(\ref{sec2:HamiltonianBosonFinal})  that in terms of fermions  
\begin{eqnarray}
 H &=& \sum_{k}\xi_{R, k} :c^+_{R, k}c_{R, k}:_F \nonumber \\
 &+& \frac1L\sum_{\bf k}\Gamma^{F, RR}_{\bf k}:c^+_{R, k_1}c^+_{R, k_2}c_{R,
k_2^\prime}c_{R, k_1^\prime}:_F \nonumber \\
  &+& \frac1L\sum_{\bf k}\Gamma^{F, RL}_{\bf k}:c^+_{R, k_1}c^+_{L, k_2}c_{L,
k_2^\prime}c_{R, k_1^\prime}:_F \nonumber\\
  &+& \frac{1}{L^2}\sum_{\bf k}\Gamma^{F, RRR}_{\bf k}:c^+_{R, k_1}c^+_{R,
k_2}c^+_{R, k_3}c_{R, k_3^\prime}c_{R, k_2^\prime}c_{R, k_1^\prime}:_F
\nonumber \\
  &+& \frac{1}{L^2}\sum_{\bf k}\Gamma^{F, RRL}_{\bf k}:c^+_{R, k_1}c^+_{R,
k_2}c^+_{L, k_3}c_{L, k_3^\prime}c_{R, k_2^\prime}c_{R,
k_1^\prime}:_F\nonumber \\
 &+& R\longleftrightarrow L +\ldots 
\end{eqnarray}
Here $\ldots$ stand for the four-fermion interactions (i.e, those involving
eight fermionic operators). 
In each of vertices $\Gamma^{F, \ldots}_{\bf k}$, we denote by  ${\bf k}$  the
vector of momenta of the fermionic operators involved. 
(As an example, ${\bf k}=(k_1, k_2, k_3, k_3^\prime, k_2^\prime, k_1^\prime)$ 
in the vertex $\Gamma^{F, RRL}_{\bf k}$.) 
To derive explicit expressions for the fermionic vertices $\Gamma^{F,
\ldots}_{\bf k}$, one substitutes the expansions 
(\ref{sec4:Fermions}) into (\ref{sec2:HamiltonianBosonFinal}) and performs
the normal ordering of resulting expressions with respect to fermionic
operators. 
In the rest of this Appendix we analyze these vertices one by one.

\subsection{Single-particle spectrum $\xi_{\eta, k}$}

The quadratic part of the fermionic Hamiltonian stems from the quadratic and the
cubic terms in the bosonic Hamiltonian (\ref{sec2:HamiltonianBosonFinal}).
For the sake of clarity, we concentrate here on the single-particle spectrum of
the right fermions. To compute $\xi_{R, k}$, we consider
\begin{eqnarray}
 H^{RR+RRR} &=& (\pi/L)\sum_{q}u_q:R_qR_{-q}:_B \nonumber \\
&+& \frac{1}{L^2}\sum_{{\bf q}} \Gamma^{B, RRR}_{\bf q}
:R_{q_1}R_{q_2}R_{q_3}:_B.
\label{app2:HRRPRRR}
\end{eqnarray}
with the densities re-expressed in term of the fermionic operators and perform
the normal ordering with respect to fermions, retaining only the contributions
quadratic in fermions.
Neglecting first the momentum dependence of $\Gamma^{B, RRR}_{\bf q}$, we get 
\begin{equation}
 \xi_{R, k}=\frac{k^2}{2 m^*}+\int_0^{q}d u_q\approx u_0 k+\frac{k^2}{2m^*}-\frac{l^2 k^3}{3}.
\end{equation}
A quick estimate shows that the contribution of the momentum-dependent terms in
the expansion of $\Gamma^{B, RRR}_{\bf q}$ is of the order $l^2 k^4/m^*$ and is
always small.

\subsection{Intrabranch two-particle vertex $\Gamma^{F, RR}_{\bf k}$}

Just as the single-particle spectrum, the coupling $\Gamma^{F, RR}_{\bf k}$
arises from the terms (\ref{app2:HRRPRRR}) of the bosonic Hamiltonian. The
contribution of the quadratic part of the bosonic Hamiltonian is easily found to
be
\begin{equation}
\Gamma^{F, RR}_{\bf k}= \frac{\pi u_0l^2}{2}(k_1-k_2)(k_1^\prime-k_2^\prime).
\end{equation}
As for the cubic coupling $\Gamma^{B, RRR}_{\bf q}$, its zero momentum part does
not contribute to $\Gamma^{F, RR}_{\bf k}$, while the contribution of its 
$O(q^2)$ terms is of the order $k^3 l^2/m^*$ and can be neglected.

\subsection{Interbranch two-particle vertices $\Gamma^{F, RL}_{\bf k}$ and
$\Gamma^{F, LR}_{\bf k}$}

\label{app:GammaFRL}

We turn now to a derivation of the fermionic vertex $\Gamma^{F, RL}_{\bf
k}$ and $\Gamma^{F, LR}_{\bf k}$ entering the two-particle interaction between
left and right sectors of the theory,
\begin{eqnarray}
 \delta H &=& \frac1L\sum_{\bf k}\Gamma^{F, RL}_{\bf k}:c^+_{R, k_1}c^+_{L,
k_2}c_{L, k_2^\prime}c_{R, k_1^\prime}:_F \nonumber \\ &+&
 \frac1L\sum_{\bf k}\Gamma^{F, LR}_{\bf k}:c^+_{L, k_1}c^+_{R, k_2}c_{R,
k_2^\prime}c_{L, k_1^\prime}:_F.
 \label{app2:deltaH}
\end{eqnarray}
Note that the terms with couplings  $\Gamma^{F, RL}_{\bf k}$ and
$\Gamma^{F, LR}_{\bf k}$ in Eq.~(\ref{app2:deltaH}) have obviously the same
structure with respect to fermionic operators, and the corresponding splitting
of the interaction is done only for notational convenience. 

The interbranch two-particle vertex $\Gamma^{F, RL}_{\bf k}$ originates from the
$\Gamma^{B, RRRL}_{\bf q}$ coupling in the Hamiltonian
(\ref{sec2:HamiltonianBosonFinal}) 
\begin{multline}
 \delta H^{B, RRRL}=
\frac{1}{L^3}\sum_{\bf q}\Gamma^{B, RRRL}_{\bf q}
:R_{q_1}R_{q_2}R_{q_3}L_{q_4}:_B\\
=\frac{6}{L^3}\sum \Gamma^{B, RRRL}_{\bf q} \Theta(q_1>q_2>q_3)R_{q_3}R_{q_2}R_{q_1}L_{q_4}.
\end{multline} 
Here we have introduced the shorthand notation
$\Theta(q_1>q_2>q_3)\equiv\Theta(q_1-q_2)\Theta(q_2-q_3)$.  
We transform now the product of three right densities into a form
normal-ordered with respect to fermions. In order to find $\Gamma^{F,RL}_{\bf
k}$, we have to collect the terms with all but one pairs of fermionic operators
replaced by the corresponding Wick contractions: 
\begin{multline}
 c^+_{Rk_3}c_{R, k_3+q_3}c^+_{Rk_2}c_{R, k_2+q_2}c^+_{Rk_1}c_{R, k_1+q_1}\longrightarrow\\
 -:c^+_{R, k_3}c_{R, k_2+q_2}:_F\langle c_{R, k_3+q_3}c^+_{R, k_1}\rangle \langle c^+_{R, k_2}c_{R, k_1+q_1}\rangle\\
 +:c^+_{R, k_3}c_{R, k_1+q_1}:_F\langle c_{R, k_3+q_3}c^+_{R, k_2}\rangle \langle c_{R, k_2+q_2}c^+_{R, k_1}\rangle\\
 -:c^+_{R, k_2}c_{R, k_3+q_3}:_F\langle c_{R, k_2+q_2}c^+_{R, k_1}\rangle \langle c^+_{R, k_3}c_{R, k_1+q_1}\rangle\\
 - :c^+_{R, k_2}c_{R, k_1+q_1}:_F\langle c_{R, k_3+q_3}c^+_{R, k_1}\rangle \langle c^+_{R, k_3}c_{R, k_2+q_2}\rangle\\
 +:c^+_{R, k_1}c_{R, k_3+q_3}:_F\langle c_{R, k_2}^+c_{R, k_1+q_1}\rangle \langle c^+_{R, k_3}c_{R, k_2+q_2}\rangle\\
 -:c^+_{R, k_1}c_{R, k_2+q_2}:_F\langle c^+_{R, k_3}c_{R, k_1+q_1}\rangle \langle c_{R, k_3+q_3}c^+_{R, k_2}\rangle.
 \end{multline}
Using the contractions of Fermi operators $\langle c_{R, k}c^+_{R,
k}\rangle=1-\langle c^+_{R, k}c_{R, k}\rangle =\Theta(k)$,   we find  
\begin{equation}
 \delta H^{B, RRRL}\rightarrow \frac{1}{L}\sum_{\bf k}\Gamma^{F, RL}_{k_1, k_2,
k_2^\prime, k_1^\prime} \left. : c^{+}_{Rk_1} c^{+}_{Lk_2}c_{L,k_2^\prime}c_{R,
k_1^\prime}:\right._F
\label{app2:DeltaHBRRRL}
\end{equation}
with 
\begin{eqnarray}
 \Gamma^{F, RL}_{\bf k} &=&
\frac{6}{L^2} \nonumber \\ &\times & \sum_{p_1, p_2>0} \Gamma^{B,
RRRL}_{-k_1+p_1+p_2, k_1+k_1^\prime-2p_1-p_2, -k_1+p_1, k_1-k_1^\prime}
 \nonumber \\
& \times &
\left[\Theta(2k_1+k_1^\prime-3p_1-2p_2)-\Theta(k_1-2p_1-p_2)\right] \nonumber \\
&+& (k_1, k_1^\prime)\rightarrow -(k_1, k_1^\prime).
\label{app2:GammaRL}
\end{eqnarray}

The behavior of the interbranch two-particle interaction at small momenta can be now inferred from Eqs. (\ref{app2:GammaRL}),  (\ref{sec2:GammaB_RRRL1}), (\ref{sec2:GammaB_RRRL2}) and (\ref{sec2:GammaB_RRRL3}).
To present the corresponding expression in a transparent form, it is convenient
to classify contributions to $\Gamma^{F, RL}_{\bf k}$ according to their
scaling with the momentum transfer between left and right movers
$Q=k_1-k_1^\prime$:
\begin{equation}
\hspace{-0.2cm}
 \Gamma^{F, RL}_{\bf k} = \frac{\pi}{m^{*2}u_0}
 \left(s_0+s_1 Q + s_2 Q^2+s_3 Q^4+s_4 Q^4\right),
 \end{equation}
 where
 \begin{eqnarray}
 s_0 &=& \alpha\left(1 -\frac{3\alpha}{2} \right)  k_1 k_1^\prime+\frac{5 (8-13
\alpha )}{32}  \alpha  l^2 k_1^2 k_1^{\prime 2},\quad\\
 s_1 &=& \frac{\alpha ^2}{8}\left[
  k_1+k_1^\prime+\frac13 l^2 k_1 k_1^{\prime}
  \right]\ln\frac{k_1^2}{k_1^{\prime 2}},\quad\\
 s_2 &=& \frac{\alpha-3\alpha^2}{6} -\frac{259 \alpha^2-114 \alpha +24}{48}  l^2
k_1 k_1^{\prime},\quad\\
 s_3 &=& \frac{1}{48} \alpha  (17 \alpha -3) (k_1+k_1^\prime)
l^2\ln\frac{k_1^2}{k_1^{\prime 2}},\quad\\
 s_4 &=& \frac{1}{576} \left(-883 \alpha ^2+276 \alpha -48\right) l^2.
\end{eqnarray}
In this work we use the amplitude $\Gamma^{F, RL}_{\bf k}$ to evaluate the
life time of the fermionic quasiparticles 
caused by triple collisions.  As we discuss in the main text,  the  momentum
transfer $Q$ between left- and right-movers in three-fermion collisions
is parametrically smaller than the typical momentum $p$ of the colliding
particles, $Q\sim p^2/u_0 m^*$. As a consequence, all but the first term
in the expansion of $\Gamma^{F, RL}_{\bf k}$ are effectively suppressed by
additional powers of mass $m^*$ in the denominator and can be neglected.
We further note that the  vertex $\Gamma^{F, RL}_{k}$ is non-singular at small
$Q$. The $1/Q$ singularity present in the bosonic vertex $\Gamma^{B, RRRL}_{\bf q}$ [the
term $q_1 q_2 q_3/q_4$ in $\tilde\tilde{\Gamma}_{\bf q}$,
Eq.(\ref{sec2:GammaB_RRRL3})] is canceled  here. We can thus neglect also 
the second term in the coefficient $s_0$ (cf. Sec. \ref{app:GammaFRRL}). We
thus obtain
 \begin{equation}
 \Gamma^{F, RL}_{\bf k}=\frac{\pi \alpha (2-3\alpha)}{2m^{*2}u_0}k_1 k_1^\prime.
 \label{app2:GammaFRLExpansionFinal}
\end{equation}
The  second contribution to the two-particle
interaction (\ref{app2:deltaH}) originates from the bosonic term $\delta
H^{RLLL}$ and can be obtained from the first one by applying the
$R\leftrightarrow L$ operation. Obviously, the corresponding vertex
$\Gamma^{LR}_{\bf k}$ is identical to $\Gamma^{RL}_{\bf k}$. 

\subsection{Interbranch three-particle vertex $\Gamma^{F, RRL}_{\bf k}$}
\label{app:GammaFRRL}

Here we derive an explicit expression for the three-particle
interbranch interaction vertex $\Gamma^{F, RRL}_{\bf k}$. Just like $\Gamma^{F,
RL}_{\bf k}$
it originates form the  bosonic interaction term $\delta H^{B, RRRL}$, Eq. (\ref{app2:DeltaHBRRRL}). 
The difference is that now we have to collect terms resulting from a
single Wick contraction in the product of the right densities,
\begin{multline}
 c^+_{Rk_3}c_{R, k_3+q_3}c^+_{Rk_2}c_{R, k_2+q_2}c^+_{Rk_1}c_{R, k_1+q_1}\longrightarrow\\
 -\langle c^+_{R, k_3}c_{R, k_2+q_2}\rangle: c^{+}_{R, k_2}c^{+}_{R, k_1}c_{R,k_1+q_1}c_{R, k_3+q_3}:_F\\
 -\langle c^+_{R, k_3}c_{R, k_1+q_1}\rangle: c^{+}_{R, k_1}c^{+}_{R, k_2}c_{R,k_2+q_2}c_{R, k_3+q_3}:_F\\
 +\langle c_{R, k_3+q_3}c^+_{R,k_2}\rangle: c^{+}_{R, k_3}c^{+}_{R, k_1}c_{R,k_1+q_1}c_{R, k_2+q_2}:_F\\
 -\langle c^+_{R,k_2}c_{R, k_1+q_1}\rangle: c^{+}_{R, k_1}c^{+}_{R, k_3}c_{R,k_3+q_3}c_{R, k_2+q_2}:_F\\
 +\langle c_{R, k_3+q_3}c^+_{R,k_1}\rangle: c^{+}_{R, k_3}c^{+}_{R, k_3}c_{R,k_2+q_2}c_{R, k_1+q_1}:_F\\
  +\langle c_{R, k_2+q_2}c^+_{R,k_1}\rangle: c^{+}_{R, k_2}c^{+}_{R,
k_3}c_{R,k_3+q_3}c_{R, k_1+q_1}:_F.
 \end{multline}
After a straightforward algebra, one finds  the interbranch three-fermion
coupling
\begin{eqnarray}
\Gamma^{F, RRL}_{\bf k} &=& -6\sign(k_2+k_1^\prime) \nonumber \\ &\times&
\sum_{p=0}^{|k_2+k_1^\prime|/2}\Gamma^{RRRL}_{k_2^\prime-k_1,
\frac{k_1^\prime-k_2}{2}+p, \frac{k_1^\prime-k_2}{2}-p,k_3^\prime-k_3}.
\nonumber \\ &&
\label{app2:GammaFRRLGeneral}
\end{eqnarray}
The result (\ref{app2:GammaFRRLGeneral}) for  $\Gamma^{F, RRL}_{\bf k}$ should
be understood as antisymmetrized with respect to incoming
($k_1$, $k_2$) and outgoing ($k_1^\prime$, $k_2^\prime$) momenta of the right
particles.

Substituting now  the small-momentum expansion (\ref{sec2:GammaB_RRRL1}),
(\ref{sec2:GammaB_RRRL2}) and  (\ref{sec2:GammaB_RRRL3}) of $\Gamma^{B,
RRRL}_{\bf q}$, we find
\begin{equation}
 \Gamma^{F, RRL}_{\bf k}=\frac{\pi^2(k_1-k_2)(k_1^\prime-k_2^\prime)}{16m^{*2}u_0}\left[\frac{s_{-1}}{Q}+s_0+\ldots
  \right],
 \label{app3:GammaFRRLFinal}
  \end{equation}
  where $Q=k_3-k_3^\prime$ is the momentum transfer between right and left
movers, and
  \begin{eqnarray}
  s_{-1} &=& 5 \alpha  l^2
\left[(k_1-k_2)^2-(k_1^\prime-k_2^\prime)^2\right],  \\
 s_0 &=& \frac{3}{2} \alpha  (13 \alpha -6) l^2
(k_1+k_1^\prime+k_2+k_2^\prime).\qquad
 \end{eqnarray}
In Eq.(\ref{app3:GammaFRRLFinal})   we dropped terms containing higher
powers of $Q$, see discussion in Appendix \ref{app:GammaFRL}.

Unlike the two-particle interaction vertex $\Gamma^{F, RL}_{\bf k}$,
the three-fermion coupling $\Gamma^{F, RRL}_{\bf k}$ is dominated by its
singular behavior at small momentum transfer $Q$ originating from the
singularity in $\Gamma^{B, RRRL}_{\bf q}$ at small $q_4$.

\section{On-shell matrix elements for triple interbranch collisions}
\label{app:MassShellAmplitudes}

The aim of this appendix is to derive the expression (\ref{sec4:T}) for the
matrix element corresponding to interbranch three-particle collisions. 

Our starting point is Eqs. (\ref{sec4:T1General}), 
(\ref{sec4:T2GeneralSimplified}), and (\ref{sec4:GammaFRRL}).
Let us consider the three incoming particles with momenta $k_1$, $k_2$ and $k_3$
and denote by $p$ and $E$ their total momentum end energy, respectively:
\begin{eqnarray}
 k_1+k_2+k_3 &=& p,\nonumber\\
 u_0 (k_1+k_2-k_3)+\frac{1}{2m^*}\left(k_1^2+k_2^2+k_3^2\right) &=& E \,.
 \label{app4:PE}
\end{eqnarray}
It is convenient to parameterize the momenta satisfying (\ref{app4:PE}) by an angle $\theta$ via
\begin{eqnarray}
 k_1 &=& -\frac{2p_F^*}{3}+\frac{p}{3}+\frac{2P_0}{3}
\left(\cos\theta+\sqrt{3}\sin\theta\right),\\
 k_2 &=& -\frac{2p_F^*}{3}+\frac{p}{3}+\frac{2P_0}{3}
\left(\cos\theta-\sqrt{3}\sin\theta\right),\\
 k_3 &=& \frac{4p_F^*}{3}+\frac{p}{3}-\frac{4}{3}P_0\cos\theta \,.
 \end{eqnarray}
  Here
 \begin{eqnarray}
 p_F^* &=& m^*u_0,\\
  P_0 &=& \sqrt{p_F^{*2}+\frac{3}{4}m^*E-\frac{p^2}{8}-\frac{p p_F^*}{4}}.
 \end{eqnarray}
 The momenta of the three outgoing particles are given by the same
expressions with the replacement $\theta\rightarrow \theta^\prime$.  
 Note that the requirement that $k_1$, $k_2$ are much smaller than $p_F^*$
restricts the angles $\theta$ and $\theta^\prime$ to 
 $|\theta|, |\theta^\prime| \lesssim  \max(k_1, k_2)/u_0 p_F^*$. 

 We now substitute the momenta parametrized by the angles $\theta$ and
$\theta^\prime$ into
Eqs.~(\ref{sec4:T2GeneralSimplified}) and (\ref{sec4:GammaFRRL}) and observe
that
 \begin{equation}
  \frac{(k1-k2)^2-(k1^\prime-k_2^\prime)^2}{k_3-k_3^\prime}=4P_0(\cos\theta+\cos\theta^\prime)\approx 8p_F^*
  \end{equation}
  and
  \begin{multline}
  \frac{k_1^2+k_2^2+4k_1k_2-k_1^{\prime2}-k_2^{\prime2}-4k_1\prime
k_2^\prime}{k_3-k_3\prime}
\\=-2p+2p_F^*-4P_0(\cos\theta+\cos\theta^\prime)\approx-4p_F^*.
 \end{multline}
The result (\ref{sec4:T}) then follows immediately.

\end{document}